\documentclass[fleqn,usenatbib]{mnras}

\usepackage{newtxtext,newtxmath}
\usepackage[T1]{fontenc}

\DeclareRobustCommand{\VAN}[3]{#2}
\let\VANthebibliography\thebibliography
\def\thebibliography{\DeclareRobustCommand{\VAN}[3]{##3}\VANthebibliography}

\usepackage{graphicx}	
\usepackage{amsmath}	
\usepackage{tablefootnote}
\usepackage{lscape} 

\newcommand{\radioexcess}{\texttt{RADIO\_EXCESS}}
\newcommand{\classsfg}{\texttt{CLASS\_SFG}}
\newcommand{\classrq}{\texttt{CLASS\_RQAGN}}
\newcommand{\classherg}{\texttt{CLASS\_HERG}}
\newcommand{\classlinelerg}{\texttt{CLASS\_LINELERG}}
\newcommand{\Nsrcs}{152\,355}

\newcommand{\Nradexs}{38\,588}
\newcommand{\NLINELERG}{12\,648}
\newcommand{\NHERG}{362}
\newcommand{\NSFG}{38\,729}
\newcommand{\NRQAGN}{18\,726}

\title[LoTSS DR2 Classifications]{The LOFAR Two Metre Sky Survey Data Release 2: Probabilistic Spectral Source Classifications and Faint Radio Source Demographics}

\author[A.~B.~Drake et al.]{A.~B.~Drake$^{1}$\thanks{E-mail: a.drake4@herts.ac.uk (ABD)},
D.~J.~B.~Smith$^{1}$,
M.~J.~Hardcastle$^1$, 
P.~N.~Best$^2$,
R.~Kondapally$^2$,
M.~I.~Arnaudova$^{1}$,
\newauthor S.~Das$^{1}$,
S.~Shenoy$^{1}$,
K.~J.~Duncan$^2$,
H.~J.~A.~R\"{o}ttgering$^{3}$,
C.~Tasse$^{4,5}$\\
$^{1}$Centre for Astrophysics Research, University of Hertfordshire, Hatfield, AL10 9AB, UK\\
$^{2}$Institute for Astronomy, University of Edinburgh, Royal Observatory, Blackford Hill, Edinburgh EH9 3HJ, UK\\
$^{3}$ Leiden Observatory, Leiden University, NL-2300 RA Leiden, Netherlands\\
$^{4}$ GEPI \& ORN, Observatoire de Paris, Universit\'{e} PSL, CNRS, 5 Place Jules Janssen, 92190 Meudon, France\\
$^{5}$ Centre for Radio Astronomy Techniques and Technologies, Department of Physics and Electronics, Rhodes University, Grahamstown, 6140, South Africa
}

\date{Accepted XXX. Received YYY; in original form ZZZ}

\pubyear{2024}

\begin{document}
\label{firstpage}
\pagerange{\pageref{firstpage}--\pageref{lastpage}}
\maketitle

\begin{abstract}
We present an analysis of 152,355 radio sources identified in the second data release of the LOFAR Two Metre Sky Survey (LoTSS-DR2) with Sloan Digital Sky Survey (SDSS) spectroscopic redshifts in the range $0.00 < z < 0.57$. Using Monte Carlo simulations we determine the reliability of each source exhibiting an excess in radio luminosity relative to that predicted from their H$\alpha$ emission, and, for a subset of 124,023 sources we combine this measurement with a full BPT analysis. Using these two independent diagnostics we determine the reliability of each source hosting a supermassive black hole of high or low Eddington-scaled accretion rate, and combine the measurements to determine the reliability of sources belonging to each of four physical classes of objects: star forming galaxies (SFGs), radio-quiet active galactic nuclei (RQAGN), and high- or low-excitation radio galaxies (HERGs or emission-line LERGs). The result is a catalogue which enables user-defined samples of radio sources with a reliability threshold suited to their science goal e.g. prioritising purity or completeness. Here we select high-confidence samples of radio sources (>90\% reliability) to report: 38,588 radio-excess AGN in the LoTSS DR2 sample (362 HERGs, and 12,648 emission-line LERGs), together with 38,729 SFGs, and 18,726 RQAGN. We validate these results through comparison to literature using independent emission-line measurements, and to widely-adopted \emph{WISE} photometric selection techniques. While our use of SDSS spectroscopy limits our current analysis to $\sim 4$ percent of the LoTSS-DR2 catalogue, our method is directly applicable to data from the forthcoming WEAVE-LOFAR survey which will obtain over a million spectra of 144\,MHz selected sources.
\end{abstract}



\section{Introduction}
Understanding the evolution of galaxies and active galactic nuclei (AGN), together with their interplay across cosmic time, remains a central focus for world-class facilities. Huge progress has been made in tracing the integrated star formation history (e.g. \citealt{HopkinsBeacom2006}, \citealt{MadauDickinson2014}), now extending to the highest redshifts ($z>6$; e.g. \citealt{Kistler2009}, \citealt{RobertsonEllis2012}, \citealt{Fudamoto2021}, \citealt{Zavala2021}, \citealt{Bouwens2022}, \citealt{Barrufet2023}, \citealt{Algera2023}), and in identifying large samples of AGN (e.g. \citealt{Mazzucchelli2017}, \citealt{Farina2022}), including some of the first supermassive black holes in the Universe (\citealt{Banados18}, \citealt{Yang20}, \citealt{Wang21}). The relative importance however of the processes of star-formation and the growth of supermassive black holes (i.e. via accretion) as function of redshift remains an open question, as indeed are the best approaches to disentangle the signatures of each process within individual galaxies. The presence of dust inside galaxies, which is a by product of the star formation process, significantly hinders our ability to trace both star formation and AGN activity, by attenuating the light sufficiently to completely obscure these processes at optical wavelengths in many galaxies. Indeed, some studies now suggest that a large fraction of the ``activity" (i.e. star formation or accretion) of the Universe occurs in this ``obscured" phase (e.g. star formation: \citealt{Magnelli2011}, \citealt{Magnelli2013}, \citealt{Gruppioni2013}, \citealt{Bouwens2016}, \citealt{Dunlop2017}, \citealt{Whitaker_2017}, \citealt{Zavala2021}; and AGN activity: \citealt{Martinez-Sansigre2005}, \citealt{Perna2015}, \citealt{HickoxAlexander2018}, \citealt{Hatcher2021}). For this reason, the radio waveband, which is impervious to dust attenuation, presents a particularly attractive solution to select samples of active extragalactic sources irrespective of their dust (or gas) content (\citealt{Condon1992}). In particular, at the lowest frequencies ($\lessapprox$ a few tens of GHz) emission from active sources is primarily in the form of synchrotron radiation -- either dominated by cosmic rays accelerated by jets, or by supernova remnants following recent star formation (at higher and lower radio luminosities respectively). \\

\noindent Assessing the so-called population mix of sources as a function of radio luminosity (i.e. the relative numbers of SFGs, AGN, and sub-classes of AGN) has been an active area of research over the past decade, as the abilities of radio observatories have rapidly evolved, leading to larger and larger samples of radio-selected extragalactic sources (e.g. \citealt{Best2012}, \citealt{Smolcic2017}, \citealt{Sabater2019}, \citealt[][]{Shimwell2017,Shimwell2019,Shimwell2022}). 

The demographics of the radio source population are a strong function of flux density; while initial surveys of the brightest sources were dominated by radio-loud AGN (e.g. \citealt{Laing1983}) with the latest surveys we are now in the regime where the source population is dominated by radio-quiet AGN and star-forming galaxies (e.g. \citealt{Best2023}, \citealt{Sabater2019}, \citealt{Das2024}). Essential to our ability to identify each of these physical classes, is the breadth and quality of complementary multiwavelength datasets that enable us to identify the optical/IR counterparts of the sources, determine their physical properties, and ultimately determine the dominant powering mechanism of each source. The gold standard technique is the use of optical spectroscopy to identify precise redshifts, and in the case of high S/N spectra, determine unambiguous source classifications thanks to the characteristic emission lines associated with star formation or AGN activity.\\

\begin{figure*}
	\includegraphics[width=\columnwidth]{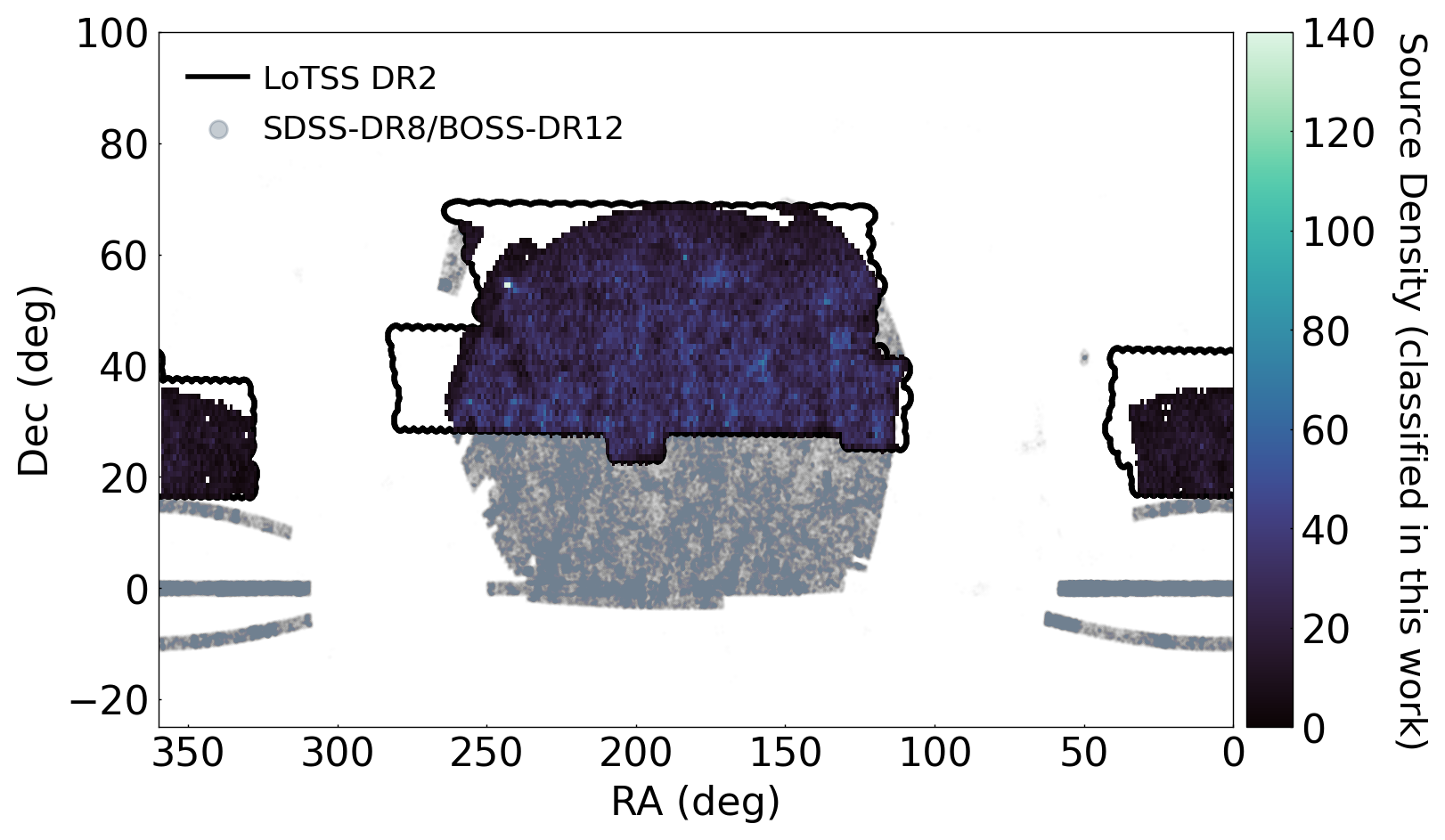}
	\includegraphics[width=\columnwidth]{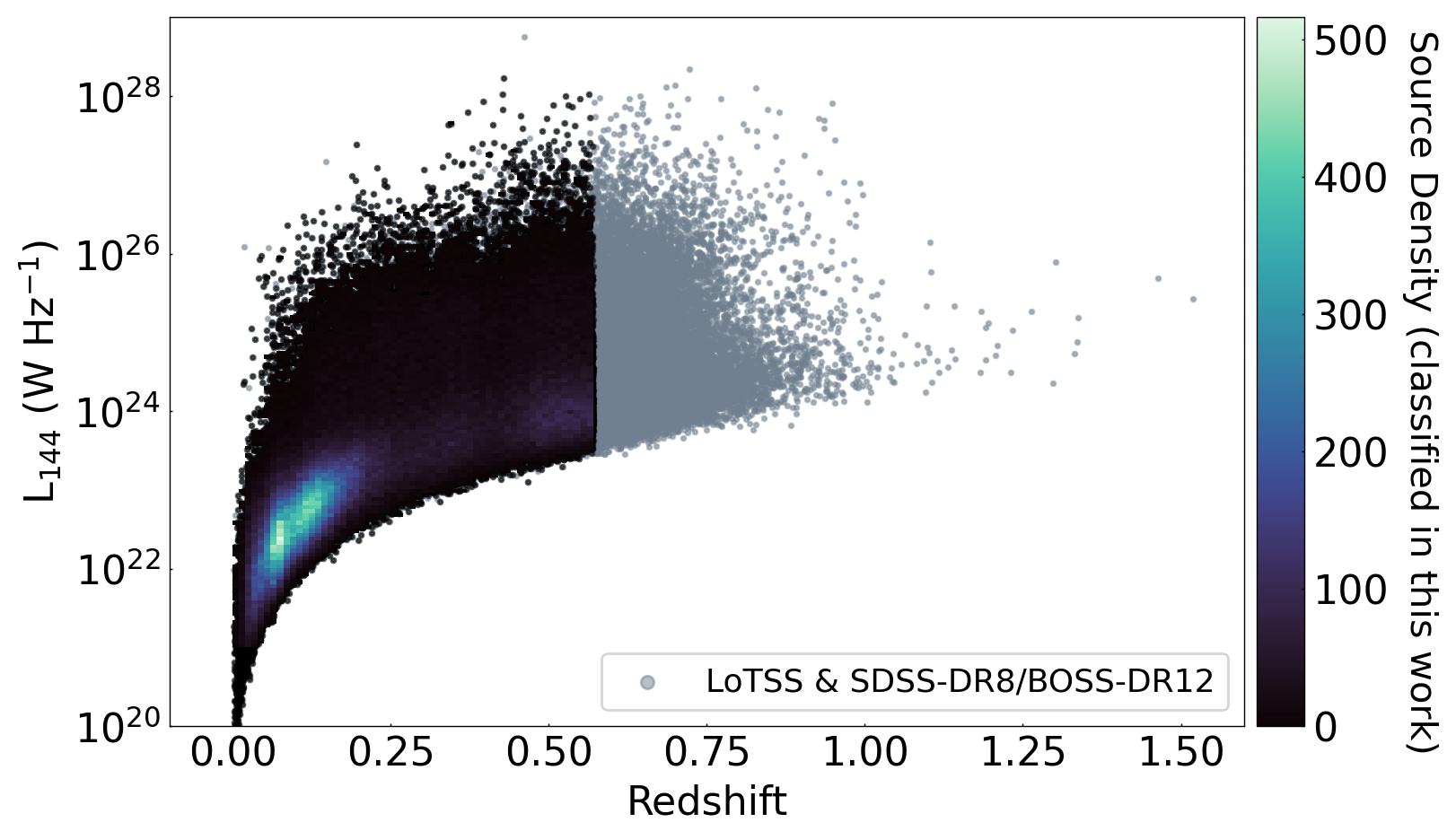}
    \caption{Left: The sky coverage of LoTSS DR2 \citep{Shimwell2019,Hardcastle2023} given by the black outline, and the Portsmouth SDSS catalogue \citep{Thomas2013} in greyscale. The objects studied in this work occupy the common region between these two surveys, and are shown as a density plot. Right: The redshift--luminosity plane, populated by sources common to LoTSS DR2 and the Portsmouth catalogues (in grey). Overlaid is a density plot to show the sources studied in this work. Note that we apply redshift cuts to ensure the H$\alpha$ spectral line is visible to the spectrograph before further analysis (see Section \ref{subsect:sample}) -- this is clearly visible at an upper redshift of $z\approx0.57$. The source density in both panels is given by the colour bars to the right.}
    \label{fig:sky_z_lum}
\end{figure*}

\noindent In a landmark work, \cite{Best2012} applied this approach to radio sources selected across $>7000$ deg$^2$ by either the NRAO (National Radio Astronomy Observatory) VLA (Very Large Array) Sky Survey (NVSS; \citealt{Condon1998}) or the Faint
Images of the Radio Sky at Twenty centimetres (FIRST) survey \citep{Becker1995}. The 1.4-GHz maps reached a completeness limit of $\sim 2.5$ mJy (NVSS; 45 arcsec beam at FWHM), and 1 mJy$/$beam (FIRST; 5 arcsec beam at FWHM). By combining this catalogue with the seventh data release of the Sloan Digital Sky Survey (SDSS) the authors were able to obtain optical spectroscopic data leading to 9,168 cross-matched sources. In this flux regime, samples of radio galaxies are dominated by AGN \citep[e.g.][]{Magliocchetti2002} and so the authors endeavoured to remove `contaminant' objects dominated by star formation from the sample following \cite{Best2005} and \cite{Kauffmann2008}, before employing a multi-stage approach to classify the remaining ``radio-AGN" (i.e. radio-excess AGN) into high- and low excitation classes (hereafter High Excitation Radio Galaxies; HERGs, and Low Excitation Radio Galaxies; LERGs). Due to the varying signal to noise in the spectra of these sources, the authors employed different techniques depending on the information available for each source, and ultimately provided a sample of 7,302 radio-excess AGN classified as either high or low excitation objects. Indeed part of the legacy of this work has been in building the consensus that supermassive black hole growth occurs via fundamentally different processes depending on the rate at which matter is accreted (see e.g. \citealt{Hardcastle2007}, \citealt{Heckman2014} and \citealt{Best2023}). At high accretion rates ($\gtrapprox 0.01$ Eddington) accretion is thought to occur via an optically-thick, geometrically-thin accretion disk (e.g. \citealt{Shakura1973}) leading to radiatively efficient accretion for which a large fraction of energy output is in the form of photons. When accretion occurs instead at a slower rate ($\lessapprox 0.01$ Eddington) the accretion flow is thought to take the form of an optically-thin, geometrically-thick accretion disk which is radiatively \emph{inefficient}. These sources are thought to be powered by advection-dominated accretion flows (ADAFs; \citealt{Narayan1995}), and emit energy mainly in kinetic form through radio jets extending into the surrounding interstellar medium. The most reliable approaches to distinguish between these sources have been explored extensively in the literature, but are frequently hampered by the inhomogeneity of multiwavelength datasets available and are rarely able to account for the often significant uncertainty in a particular classification.\\

 \noindent A subsequent major leap forward came from the VLA COSMOS 3 GHz Large Project \citep{Smolcic2017} which provided the state of the art deep-field work reaching 2 orders of magnitude deeper than \cite{Best2012} over a 2.6 square degree map with a mean RMS of 2.3\,$\mu$Jy\,beam$^{-1}$. Using exquisite panchromatic photometric observations available in the COSMOS field, \cite{Smolcic2017} analysed the population mix of 9,161 radio sources with multiwavelength counterparts. The authors used a variety of criteria to separate SFGs and AGN, and further classify the AGN into `low-to-moderate' and `moderate-to-high' luminosity, as analogues of the high- and low-excitation classes outlined in \cite{Best2012}. They report that at flux levels below $\sim 100 \mu$Jy at 3GHz, the SFG population begins to dominate the source counts, and use the demographic breakdown of their sources to predict that future surveys such as those carried out using the SKA will provide an efficient manner in which to select large samples of SFGs.\\
 
 \noindent The SKA-pathfinder ``MeerKAT" for instance has recently conducted the MeerKAT International GHz Tiered Extragalactic Exploration (MIGHTEE) Survey, and in early re-analysis of the COSMOS field, \cite{Whittam2022} find already 54\% of 5,223 radio sources classified in their work are likely to be SFGs, becoming the dominant population below 1.4GHz flux densities of $\approx 150 \mu$Jy.  \\
    
\noindent Radio surveys now provide a means to detect vast numbers of extragalactic sources, for instance; The International LOFAR Telescope (hereafter ``LOFAR"; \citealt{vanHaarlem2013}), which, operating at frequencies below 200\,MHz, is sensitive to a broader range of timescales than higher-frequency work, is also able to target synchrotron radiation almost entirely uncontaminated with thermal emission. LOFAR has achieved a sensitivity and survey speed which has enabled the LOFAR Two-Metre Sky Survey (LoTSS) - aiming to provide coverage of the entire nothern sky at a central frequency of 144 MHz, reaching $\sim 100 \mu$Jy/beam,  with a resolution of $6$ arcseconds at FWHM. LoTSS also includes a deep tier aiming to achieve depths of $10-15 \mu$Jy beam$^{-1}$, across 25 square degrees in total. The LOFAR Deep-Fields (ELAIS-N1, Lockman Hole and Bootes) are complemented with the deepest multiwavelength datasets available in the northern sky, enabling panchromatic SED fitting to identify the optical counterparts to the radio sources, and infer their properties. As a consensus on the physical picture of AGN accretion modes has emerged in the literature (e.g. \citealt{Best2014}, \citealt{Hardcastle18}, \citealt{HardcastleCroston20}) it has become prudent to search for AGN signatures via different approaches designed to detect characteristics of the two different modes. Indeed the signature of these fast- and slow-accretors manifests very differently in the observable properties of sources (although also see \citealt{Whittam2022} and Kondapally \textit{in prep}, for a discussion of sources potentially occupying a continuum of states between the fast and slow modes). In recent work, cite{Best2023} and \cite{Das2024} apply SED-fitting codes to photometry of LOFAR-detected samples in the LoTSS deep fields, and aim to identify AGN thought to be rapidly accreting (characterised by a hot, radiatively-efficient accretion disk inside a dusty torus) in the SED fits, before searching independently for the likely more slowly accreting (i.e. thought to be radiatively inefficient) AGN via an excess of radio emission relative to that predicted from their SFRs. Via this method \cite{Best2023} showed that 144MHz flux densities (F$_{{\rm{144}}}$) below 1 mJy are dominated by SFGs, comprising 90 \% of the population at 100 $\mu$Jy. \\

\noindent  In contrast to the depth of the LoTSS Deep Fields, the wide LoTSS tier is complemented with far shallower photometry, but benefits from high quality spectroscopic sky surveys; e.g. SDSS \citep{York2000}. At present 4\% of LoTSS-wide sources in the second data release (DR2; \citealt{Shimwell2022}) have been observed with the SDSS or BOSS spectrographs. The diagnostic power of the full LoTSS dataset ($> 4$ million sources in the LoTSS DR2 catalogue; \citealt{Hardcastle2023}) will be dramatically increased as the spectroscopic information available for these sources becomes more complete. A dramatic increase in the optical spectroscopic follow-up of LoTSS DR2 will be provided by the WEAVE-LOFAR survey \citep{Smith2016}, using the William Herschel Telescope's (WHT) new massively multiplexed optical spectrograph the WHT Enhanced Area Velocity Explorer (WEAVE; \citealt{Dalton2012}), which is currently  in the final stages of commissioning. At the sensitivity now routinely achieved with LOFAR, the wide tier of the LoTSS surveys is sensitive to a flux range and volume likely to contain every physical class of extragalactic object (radio-loud AGN, radio-quiet AGN, and SFGs), making several aspects of source classification of particular importance. Firstly, our ability to classify sources with a homogeneous method which can be applied to the full sky, secondly, providing \emph{secure} classifications (for which the gold-standard is accompanying optical spectra), together, finally, with a measure of reliability of a particular classification. Bearing in mind the imminent and voluminous dataset from WEAVE-LOFAR and from other wide-area spectroscopic surveys such as that from the Dark Energy Spectroscopic Instrument (DESI), it is important to develop tools now to cleanly identify sources with a homogeneous method which can be applied to large optical spectroscopic datasets coupled with sensitive low-frequency radio data. In addition, various science goals require the selection of samples of objects in a particular class in different ways, for instance prioritising purity or completeness, and thus a probabilistic element to classifications (i.e. the reliability of given source belonging to particular class of objects) offers a well defined way to examine different subsets of the population. \\

\noindent In this work, we develop and validate a probabilistic spectroscopic classification scheme with which to homogeneously classify large samples of radio-selected sources, and derive the source demographics of sources from the LoTSS DR2 wide tier using optical counterpart associations from \cite{Hardcastle2023} and existing SDSS spectra processed and presented in the Portsmouth catalogue \citep{Thomas2013}. The paper proceeds as follows: in Section \ref{Sect: Data} we describe the dataset generated by LOFAR and the subset of objects observed with SDSS and analysed by Portsmouth, in Section \ref{Sect: Method} we present our classification scheme before validating the method in Section \ref{Sect: Validation}. In Section \ref{sec:results} we present the demographics of the faint radio source population from LOFAR, and compare to previous studies, finally in Section \ref{Sect: Concl} we summarise our results. Throughout this work we assume a $\Lambda$CDM cosmology with $\Omega_{\rm{m}}$ = 0.3, $\Omega_{\rm{\Lambda}}$ = 0.7 and H$_0$ = 70 km s$^{-1}$ Mpc$^{-1}$. \\

\section{Data}
\label{Sect: Data}

\subsection{LoTSS}
The LOFAR Two Metre Sky Survey \citep[LoTSS;][]{Shimwell2017,Shimwell2019,Shimwell2022} is the flagship extragalactic survey conducted by the LOFAR International Telescope \citep{vanHaarlem2013} using the high-band antennas (HBA) at 120–168 MHz with a central frequency of 144MHz. LoTSS aims to survey the entire northern sky reaching average sensitivities of $\approx\!\!100\,\mu$Jy at most declinations, with a resolution of 6 arcseconds using Dutch stations alone. The second data release from LoTSS (DR2; \citealt{Shimwell2022}) consists of 841 LOFAR pointings, and covers $5,700$\,deg$^2$ of the extragalactic sky ($\approx\!27$ per cent) focused on areas of high Galactic latitude. DR2 is currently the largest radio catalogue in existence, containing 4.2 million sources at $\ge 5\sigma$ significance. The process of assigning optical counterparts to this catalogue is described in \citet{Hardcastle2023}. In brief, the likelihood-ratio cross-matching method developed for DR1 is applied \citep{Williams2019} following approximately the same decision tree to flag sources for which additional input via the citizen science project ``Radio Galaxy Zoo (LOFAR)" is required. Any deviations from the workflow described in \cite{Williams2019}
are highlighted in section 4 of \citet{Hardcastle2023}. At the conclusion of this process, optical counterparts are identified for 85 per cent of sources in LoTSS DR2 and 58 per cent have a good redshift estimate from existing SDSS spectroscopy, or from photometric redshifts \citep{Duncan2021}.  

\subsection{Existing spectra from SDSS and BOSS}
Optical spectroscopic information for a subset of LoTSS sources is available from the Sloan Digital Sky Survey (SDSS; \citealt{York2000}). To enable a cross match to the largest number of sources possible we make use of two data releases - SDSS-DR8 \citep{Aihara2011} and BOSS-DR12 \citep{Eisenstein2011}. Based on these catalogues, \cite{Thomas2013} provide line flux measurements using the fitting-code ``{\sc{gandalf}}" \citep{Sarzi2017} of the major optical emission lines typically used for the identification of radiative AGN: H$\beta$$_{\rm{4861}}$, [O{\sc{iii}}]$_{\rm{5007}}$, H$\alpha$$_{\rm{6563}}$ and [N{{\sc{ii}}}]$_{\rm{6584}}$ (hereafter H$\beta$, [O{\sc{iii}}], H$\alpha$ and [N{{\sc{ii}}}]) together with their associated errors, and fit warnings. In addition we make use of the value-added MPA-JHU catalogue \citep{Brinchmann2004}\footnote{https://www.sdss4.org/dr17/spectro/galaxy\_mpajhu/} primarily for the purpose of examining the derived physical properties of a subset of the objects we classify, specifically their star formation rates (SFRs) and stellar masses based on the methods of \citealt{Kauffmann2003} and \citealt{Tremonti2004}. 

\subsection{Sample Selection}
\label{subsect:sample}
The parent sample is the catalogue of 4.2 (3.6) million radio sources in the LoTSS DR2 (cross-identified) catalogue which is described in detail by \cite{Hardcastle2023}. Figure \ref{fig:sky_z_lum} shows the distribution on sky of the LoTSS DR2 sources (black outline) in comparison to SDSS-DR8\slash BOSS-DR12 -observed objects in the Portsmouth spectroscopic catalogue \citep[][shown in blue]{Thomas2013}. The dark blue area shows the intersection of LoTSS DR2 and Portsmouth catalogues. To associate the spectroscopic information with the low-frequency radio data, we perform a positional cross match between the positions of the optical counterparts identified for the 144\,MHz sources in \citet{Hardcastle2023}, and those in the \citet{Thomas2013} catalogue, using a nearest-neighbour algorithm with a maximum search radius of 1 arcsec. This gives us a sample of 208,816 LOFAR sources with SDSS spectroscopic information, representing 5.7 per cent of the 3,672,413 LOFAR sources that fall within the region covered by the Portsmouth SDSS DR8\slash DR12 catalogues. 

We then apply a series of cuts before proceeding to classify sources, the numbers of sources removed at each stage are recorded in Table \ref{tab:cuts}. Firstly, as a key part of our workflow (discussed in Section \ref{Sect: Method}) compares the H$\alpha$ luminosity (L$_{\rm{{H\alpha}}}$) with the 144\,MHz luminosity (L$_{{\rm{144}}}$), we remove sources which fall at redshifts where H$\alpha$ is not visible to the SDSS and SDSS-BOSS spectrographs. For SDSS DR8\footnote{https://classic.sdss.org/dr1/products/spectra/index.php} the red end of the spectrograph cuts off at 9,200\,\AA\ giving an upper redshift cut of $z = 0.385$. Likewise, for the BOSS\footnote{https://www.sdss.org/instruments/boss-spectrographs/} subset, spectrograph transmission drops off at 10,400\,\AA, and we impose an upper redshift cut at $z = 0.570$. We next remove any remaining sources for which the reported H$\alpha$ emission line measurement is unreliable (i.e. a flux and uncertainty equal to zero in the Portsmouth catalogue, or a reported H$\alpha$ flux that is infinite or entirely missing). Examining the redshift-log$_{\rm10}($L$_{\rm{{H\alpha}}}$) plane reveals several objects with unphysical Balmer line luminosities; to remove these objects, we consider only those with $4 < \log_{\rm10}$(L$_{\rm{{H\alpha}}}\slash L_\odot) < 13$.

Finally, visual inspection reveals multiple sources that are selected as high-probability radio excess but are actually nearby SFGs. This arises because the H$\alpha$ measurement is made in the 3-arcsec SDSS fibre aperture (i.e. containing only the centre of the galaxy) while the catalogue radio flux is representative of the total. To limit the influence of this effect, we consider the distribution of the effective radius, $r_{50}$, from the Legacy Surveys imaging as a function of redshift; and choose to remove sources with $r_{50} > 15$ arcsec. It may be possible to improve upon this with more sophisticated modelling, however we defer that analysis for a future investigation.

The final catalogue contains 152 355 sources, within which we define one further subset - the objects for which we can apply full line-ratio analysis to determine a physical classification. To do so we exclude all objects for which any of the principal BPT diagnostic emission lines (H$\alpha$, H$\beta$, [O{\sc{iii}}] and [N{\sc{ii}}]) are recorded to have a flux and associated error equal to zero in the Portsmouth catalogue (of course the requirement is already imposed on the H$\alpha$ line in the previous step). This subset contains 124,023 sources with spectral measurements in all four of the BPT lines. Each of these pre-processing steps are recorded in Table \ref{tab:cuts}, and show how many sources are dropped at each stage from the initial match between the LOFAR DR2 catalogue \citep{Hardcastle2023} and the SDSS spectra \citep{Thomas2013}.

\begin{table}
    \caption{A summary of the pre-processing steps applied to the initial matched catalogue combining  ~\protect\cite{Hardcastle2023} and SDSS-DR8 and BOSS-DR12 ~\protect\citep{Thomas2013}. The first column describes the cut applied (as discussed in detail in Section \ref{subsect:sample}), the central and right-hand columns list how many sources were removed in each step, and how many remain in the catalogue afterwards. In the bottom two lines we re-iterate the final numbers of sources in the full radio-excess catalogue, and the fully classified catalogue.}
    \centering
    \begin{tabular}{l|c|c}
      \hline
		\hline
        {\bf{Pre-processing Step}} & {\bf{Sources Removed}} & {\bf{Sources Remaining}} \\
        \hline
        \hline
        Initial match & -- & 208 816 \\
         Redshift cuts & 34 956 & 173 860 \\
         fH$\alpha$ \& eH$\alpha$ == 0.0 & 50 826 & 154 145 \\
         Infinite/missing H$\alpha$ & 101 & 154 044 \\
        redshift-H$\alpha$ plane & 168 & 153 876 \\
        R50 cut & 1521 & 152 355 \\
        \hline
         Missing BPT line & 28 332 & 124 023 \\
         \hline
        Final Radio-Excess Catalogue && 152 355 \\
        Final Fully-Classified Catalogue && 124 023 \\
                \hline
        \hline
    \end{tabular}
    \label{tab:cuts}
\end{table}

\begin{figure}
	\centering
 \includegraphics[width=\columnwidth]{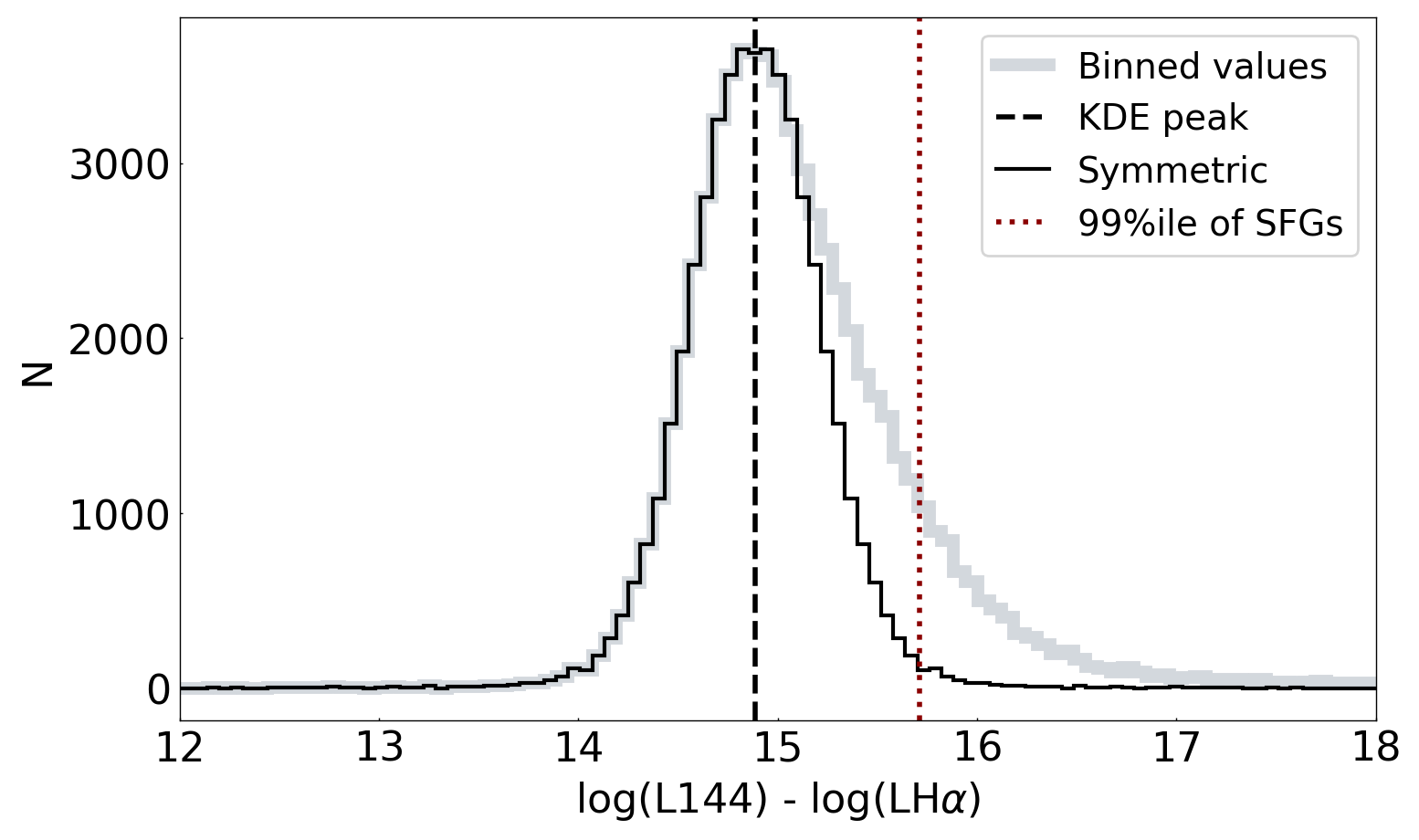}
    \caption{Histogram showing the distribution in $\log_{10} L_{144} - \log_{10} L_{\mathrm{H}\alpha}^\mathrm{corr}$ of sources with $>5\sigma$ detections in both LoTSS and their H$\alpha$ emission line (shown in light grey). Overlaid is the KDE estimate of the modal value, indicated by the vertical dashed line. The estimated distribution for SFGs is shown in black, obtained by symmetrizing the left-hand side of the distribution about the KDE peak, and also overlaid is the 99$^\mathrm{th}$ percentile of the inferred SFG distribution, above which we classify objects as having a 144\,MHz excess (red dotted line).}
    \label{fig:excess_line}
\end{figure}

\begin{figure*}
        \includegraphics[width=\columnwidth]{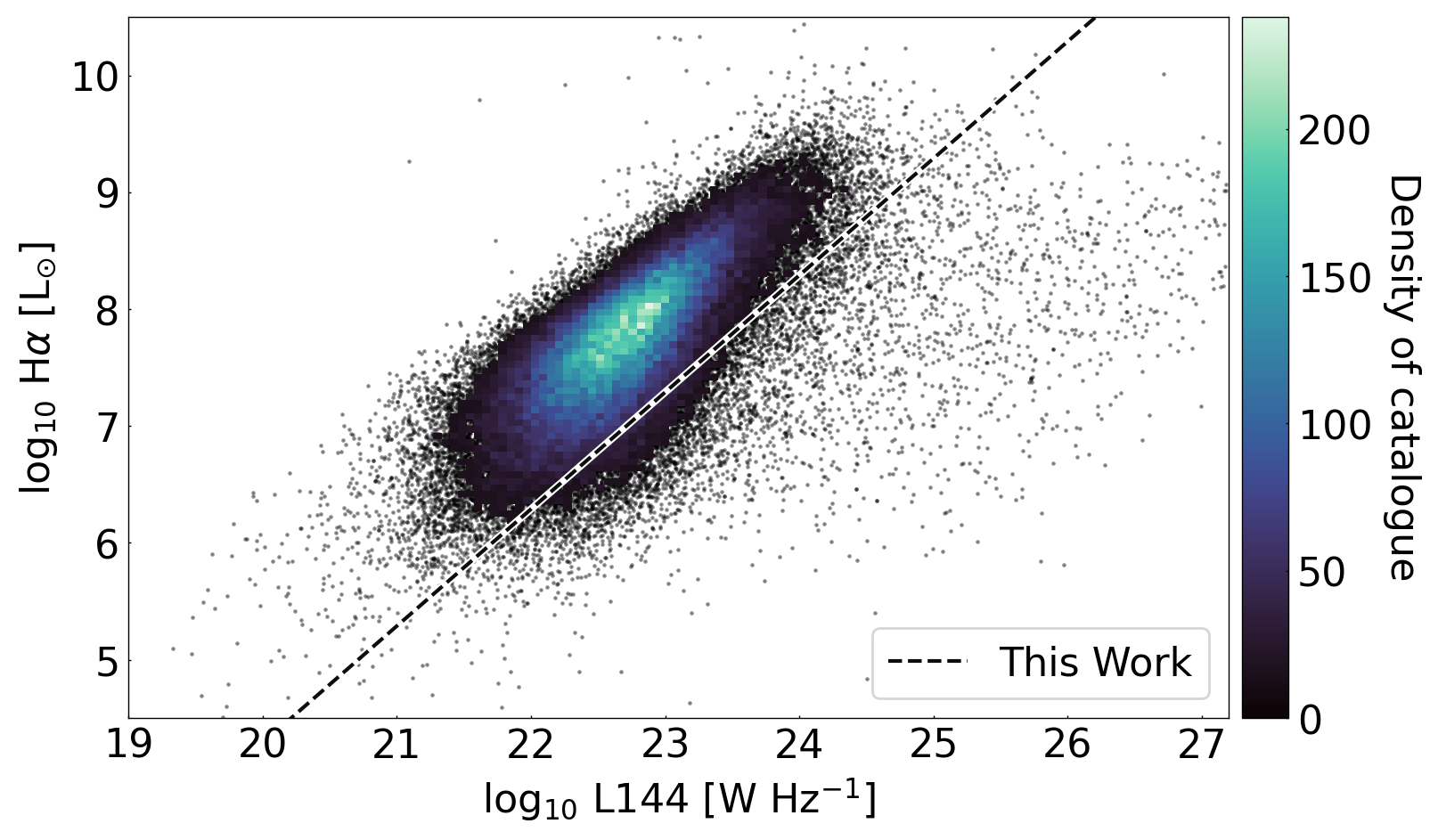}
        \includegraphics[width=\columnwidth]{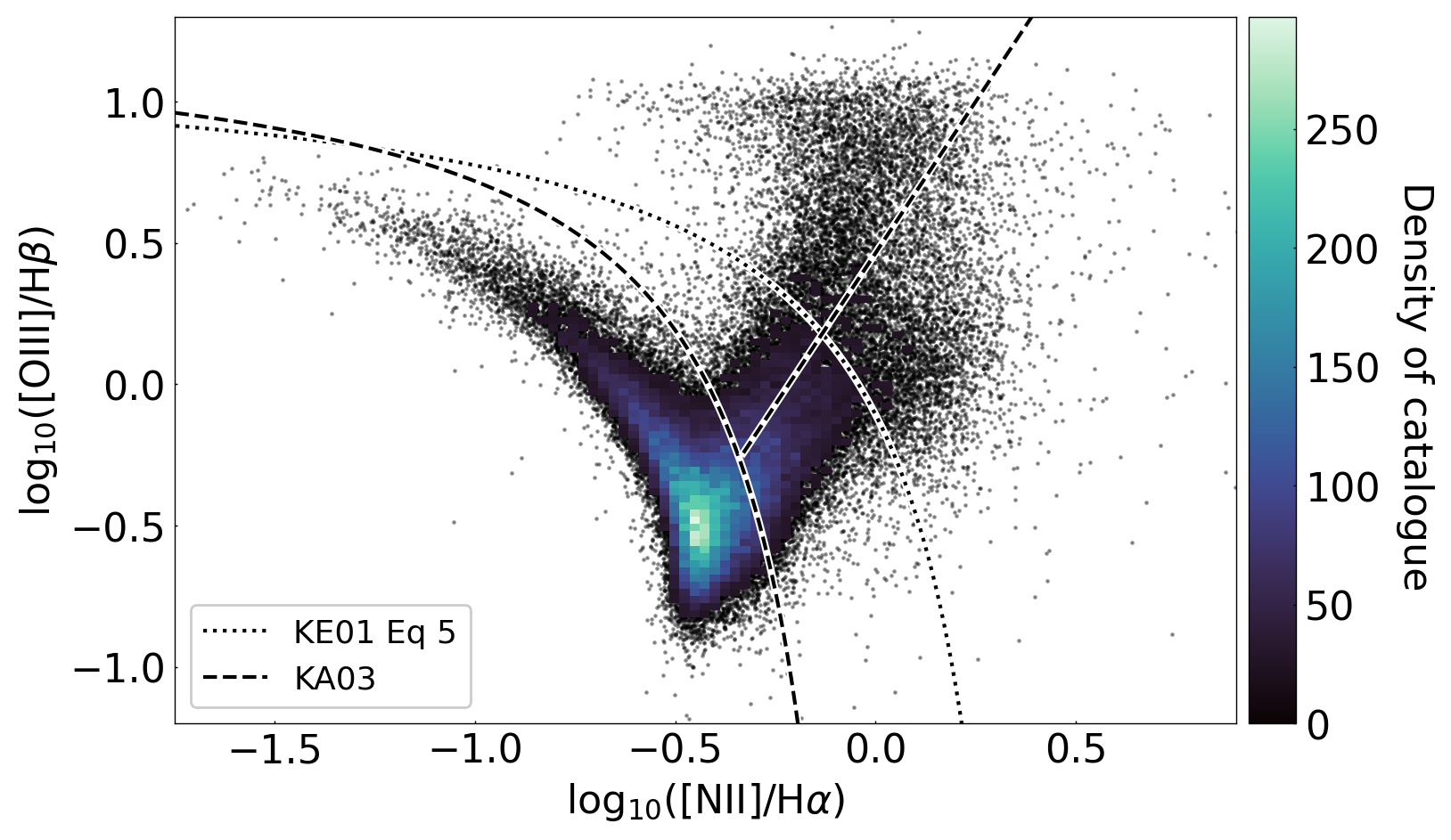}	
    \caption{The two diagnostic diagrams used for our probabilistic source classification. Left: the Balmer-decrement corrected H$\alpha$ luminosity compared to the 144\,MHz luminosity. This plot is used to identify those (realisations of) sources exhibiting an excess in radio luminosity above what might be expected on the basis of the H$\alpha$ luminosity (i.e. star formation in the host galaxy). Right: the BPT diagram \citep{BPT} used to identify those (realisations of) sources which contain radiative AGN, and distinguish between high- and low-excitation sources for those objects with recorded fluxes in all four diagnostic lines (at any S/N ratio). See text for details.}
    \label{fig:excess_and_bpt}
\end{figure*}

\begin{figure}
    \includegraphics[width=\columnwidth]{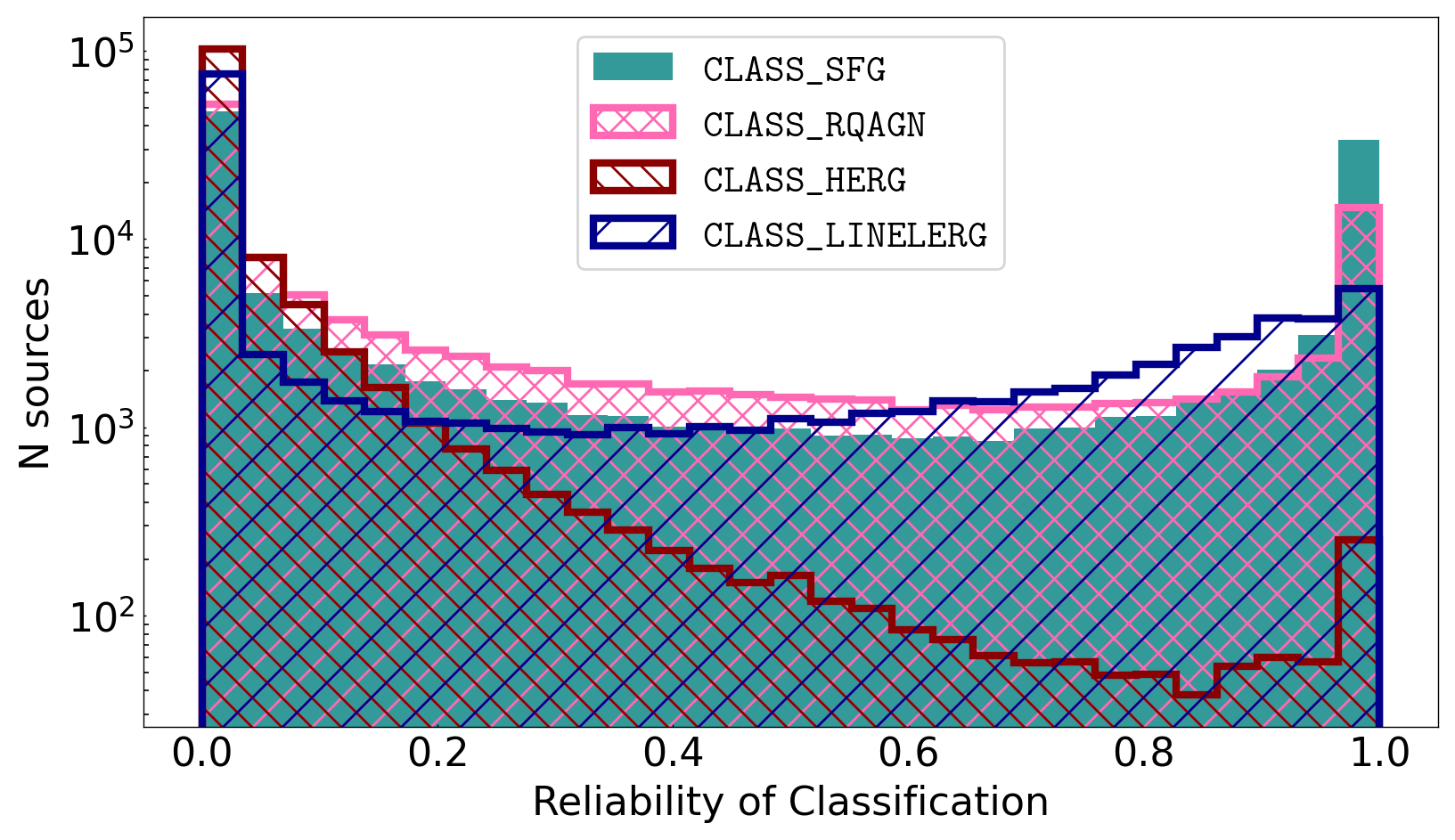}
    \caption{The distribution of \classsfg, \classrq, \classherg\ and \classlinelerg\ for our sample, shown as teal shading, pink, red and blue hatching, respectively, as indicated by the legend. }
    \label{fig:probs_by_class}
\end{figure}

\section{Classification Method}
\label{Sect: Method}

The sample of extragalactic radio sources detected by LOFAR in the Wide LoTSS tier reaches radio flux densities which previous works have shown to be a mixture of star forming galaxies and low-level AGN activity (e.g. \citealt{Best2023}, \citealt{Das2024}). As discussed above, coupled with the volume observed in LoTSS DR2 alone, LoTSS-Wide is sensitive to the full range of physical classes of objects. The ability to separate SFGs from AGN-dominated sources as well as distinguish between modes of AGN activity thus becomes increasingly critical. Although much has now been achieved in the LoTSS Deep-Fields using photometric data, we aim here to be able to identify the signature of an AGN in any given source via either accretion mode using optical spectroscopic data from SDSS (rather than enacting a search for a particular class of object and discounting other sources). We search first for radio excess relative to star formation activity, and then consider the emission line properties of the sources via the BPT diagram for objects exhibiting all four diagnostic lines. In addition to providing ``maximum likelihood" classifications, this method allows us to quantify the doubt associated with a given classification, though a Monte-Carlo simulation in which we generate 1000 realisations of our matched catalogue and combine the resultant classifications of the individual catalogues into an overall probabilistic set of of classifications. We describe each stage of this process in the following subsections. \\

\subsection{Identification of sources with a radio excess}
\label{subsect:rad_excess}
The relationship between SFR and radio luminosity arising from synchrotron emission generated in supernova remnants (shocks) is a well-studied and calibrated phenomenon (e.g. \citealt{Bell2003}) including investigation with low frequency radio data (e.g. \citealt{Gurkan2018}, \citealt{Smith2021}). This relationship offers us a means through which to predict the level of radio emission associated with star formation for a particular source given a reliable SFR estimate. Sources with a radio luminosity in excess of the level predicted for their SFR can then be flagged as exhibiting a radio excess inferred to arise from nuclear activity (i.e. an AGN). The most appropriate way to quantify this excess has been investigated in the literature by several authors e.g. \cite{Sabater2019}. Here, we compare the radio luminosity at 144 MHz (L$_{{\rm{144}}}$) with the H$\alpha$ luminosity in solar units corrected for extinction at H$\alpha$ via the Balmer decrement (i.e. see left-hand panel of Figure \ref{fig:excess_and_bpt}). To do this we calculate values of extinction, A$_v$, for all sources with both the H$\alpha$ and H$\beta$ lines recorded in the Portsmouth catalogue, and apply the standard Balmer correction for all physical (positive) values of A$_v$ (in the case of a negative value of A$_v$, which is unphysical, no extinction correction is applied). For those objects without an H$\beta$ measurement, we instead apply a statistical extinction correction using the distribution of A$_v$ values for all objects exhibiting a $5\sigma$ detection in both the H$\alpha$ and H$\beta$ lines. For each of our realisations, we then draw an extinction correction randomly from this distribution for each object without an H$\beta$ measurement. A similar method was also implemented by \cite{Best2012}, and \cite{Sabater2019} on the first data release of LoTSS in the HETDEX Spring Field \citep{Hill2008}. Figure \ref{fig:excess_line} presents a histogram of the ratios of radio to H$\alpha$ luminosities of sources with $>5\sigma$ detections in both LoTSS and the H$\alpha$ emission line (plotted as $\log_{10} L_{144} - \log_{10} L_{\mathrm{H}\alpha}^\mathrm{corr}$). By approximating the SFG population as a Gaussian distribution symmetric about the peak  estimated from the kernel density distribution (i.e. mirroring the left-hand side of the distribution around the KDE peak) we can infer the 99th percentile of the SFG distribution, and define objects lying to the right of this as having a radio excess. This means that at the position of the division line 99\% of SFGs will lie to the left in ($\log_{10} L_{144} - \log_{10} L_{\mathrm{H}\alpha}^\mathrm{corr}$). Sources to the right of the line are assumed to exhibit a radio-excess due to nuclear activity. Identification of a radio excess in isolation, does not have implications for the accretion rate or mode of the AGN; i.e. sources with both high- and low- Eddington-scaled accretion rates (radiatively efficient or inefficient AGN) can show a radio excess relative to their SFRs. 

We define sources as having a radio excess if:

\begin{equation}
    \log_{10} (L_\mathrm{144\,MHz} / \mathrm{W\,Hz}^{-1})  > \log_{10} (\rm{L}_{\mathrm{H}\alpha}^{Corrected}/ L_\odot) + 15.71
    \label{eq:excess}
\end{equation}

\subsection{Identification of radiative AGN}
\label{subsect:radiative_agn}

The identification of radiatively-efficient AGN has been attempted and refined via various different methods depending on the data available. When spectroscopic data are available the presence of this accretion disk can be inferred via optical emission-line ratios, most frequently via the so-called `BPT' diagram \citep{BPT}. Ionisation by the hot material in the inner-parts of the accretion disk excites species of atoms with higher ionisation energies than possible from ionising stellar radiation alone, altering the ratios of pairs of emission lines of different species (pairs of lines are selected at similar wavelengths such that differential dust extinction along the spectrum is not an issue). The demarcation between objects powered purely by star formation, and those dominated by light from an accretion disk has been investigated thoroughly by multiple authors, with one of two divisions mainly being adopted depending on the science question at hand. In this work we adopt the \cite{Kauffmann2003} division line (which is empirically derived to trace the upper boundary of the SFG locus in the N[{\sc{ii}}]/H$\alpha$ -- [O{\sc{iii}}]/H$\beta$ plane) to class objects as radiative AGN (see Figure \ref{fig:excess_and_bpt} right-hand panel). We also consider the \cite{Kewley2001} maximal starburst line which uses a combination of stellar population synthesis \citep{Fioc1997} and photoionisation models \citep{Sutherland1993} to derive a theoretical upper limit for galaxies purely driven by star formation. Sources with emission line ratios robustly measured to lie beyond this line definitively require the presence of a radiative AGN in order to reproduce the source's position on the BPT diagram, whereas those lying between the two divisions are considered to be `composite' sources, with spectra likely to exhibit signs of both star formation and radiative AGN activity. An additional separation can be made between objects with hard-ionising spectra characteristic of Seyfert galaxies \citep[e.g.][]{Weedman1977}, and those characteristic of lower ionisation radiative AGN - Low Ionisation Nuclear Emission Regions (LINERs; e.g. \citealt{Heckman1980}, \citealt{Cid2011}). This means every source with all four BPT emission lines measured will fall into one of three classes - \texttt{BPT\_SFG}, \texttt{BPT\_SEY} (Seyfert), or \texttt{BPT\_LIN} (LINER). Some fraction of \texttt{BPT\_SEY} and \texttt{BPT\_LIN} sources will fall in the composite region between the \cite{Kauffmann2003} and \cite{Kewley2001} demarcation lines. Thus we also report three subsets of the above categories \texttt{BPT\_COMP} (subset of \texttt{BPT\_SEY} and \texttt{BPT\_LIN} which fall in the composite region), and further divide \texttt{BPT\_COMP} into those sources on the Seyfert and LINER sides of the division line proposed in \cite{Kauffmann2003}; \texttt{BPT\_CSEY} and \texttt{BPT\_CLIN} respectively.

\begin{table*}
    \centering
    \caption{Column headings included in the table of classifications we provide with this work. In the first column we list the column heading, in the second column a brief description of its content, and in the final column we describe the values the column can take.}
    \begin{tabular}{p{0.15\linewidth}p{0.6\linewidth}p{0.2\linewidth}}   	
        \hline
		\hline
		{\bf{Column Heading}} & {\bf{Description}} & {\bf{Values}}\\
		\hline
		\hline
        \texttt{CLASS\_z\_WARNING} & Redshift warning:  ($z_\mathrm{best} - z_\mathrm{Ports}) >0.001$ & Binary 0/1 (1 = warning) \\
        \texttt{RADIO\_EXCESS} & Fraction of realisations in which L$_\mathrm{144\,MHz}$ is larger than predicted values for 99\% of SFGs (using Balmer-corrected L$_{\rm{H\alpha}}$) & Decimal between 0-1 \\
        \texttt{BALMER\_CORR\_WARNING} & Balmer correction derived without H$\beta$ detection. Extinction is instead drawn randomly from the distribution of values calculated for sources with $5\sigma$ H$\alpha$ and H$\beta$ detections.  & Binary 0/1 (1 = warning) \\
        \texttt{BPT\_SFG} & Reliability of line ratios lying in SFG area of the BPT diagram{{\color{blue}{$^1$}}} &  Decimal between 0-1 \\
        \texttt{BPT\_SEY} & Reliability of line ratios lying in Seyfert area of the BPT diagram{{\color{blue}{$^1$}}} & Decimal between 0-1\\
        \texttt{BPT\_LIN} & Reliability of line ratios lying in LINER area of the BPT diagram{{\color{blue}{$^1$}}}& Decimal between 0-1\\
        \texttt{BPT\_COMP} & Reliability of line ratios lying in Composite area of the BPT diagram{{\color{blue}{$^{1,2}$}}}& Decimal between 0-1\\
        \texttt{BPT\_CSEY} & Reliability of line ratios in Composite-Seyfert area of the BPT diagram{{\color{blue}{$^{1,3}$}}}& Decimal between 0-1\\
        \texttt{BPT\_CLIN} & Reliability of line ratios in Composite-LINER area of the BPT diagram{{\color{blue}{$^{1,3}$}}} & Decimal between 0-1\\
        \texttt{BPT\_ML} & Maximum likelihood BPT result (i.e. using Portsmouth catalogue fluxes)& s, S, L, C, X, Y{{\color{blue}{$^4$}}} \\
        \texttt{zscore} & Significance of deviation from the null hypothesis that maximum likelihood classification is correct & Decimal\\
        \hline
        \texttt{CLASS\_SFG} & Reliability of \texttt{CLASS\_SFG} classification - \texttt{CLASS\_SFG} requires no radio excess to have been measured, and a BPT line ratio falling in the SFG part of the diagram & Decimal between 0-1\\
        \texttt{CLASS\_RQAGN} & Reliability of \texttt{CLASS\_RQAGN} classification - \texttt{CLASS\_RQAGN} requires that no radio excess is measured, and a BPT line ratio above the \cite{Kauffmann2003} demarcation line & Decimal between 0-1\\
        \classherg & Reliability of \texttt{CLASS\_HERG} classification - \texttt{CLASS\_HERG} requires that a radio excess \emph{is} measured, and that the BPT line ratio lies in the Seyfert region of the plot & Decimal between 0-1\\
        \classlinelerg & Reliability of \texttt{CLASS\_LINELERG} classification - \texttt{CLASS\_LINELERG} requires that a radio excess \emph{is} measured, and that the BPT line ratio lies in either the SFG or LINER regions & Decimal between 0-1\\

		\hline
		\hline
  \multicolumn{3}{l}{\footnotesize{{{\color{blue}{$^1$}}}Reliability calculated as fraction of catalogue realisations falling in this class}} \\ 
  \multicolumn{3}{l}{\footnotesize{{{\color{blue}{$^2$}}}\texttt{BPT\_COMP} is the sum of the subsets \texttt{BPT\_SEY} and \texttt{BPT\_LIN} which fall between the \cite{Kauffmann2003} and \cite{Kewley2001} demarcation lines}}\\
   \multicolumn{3}{l}{\footnotesize{{{\color{blue}{$^3$}}}\texttt{BPT\_CSEY} and \texttt{BPT\_CLIN} classes occupy the composite area of the BPT diagram, and fall to the left and right of the Seyfert LINER division, respectively}}\\

   \multicolumn{3}{l}{\footnotesize{{{\color{blue}{$^4$}}}s=SFG, S=Seyfert, L=LINER, C=Composite, X=composite-Seyfert, Y=composite-LINER}}\\
\end{tabular}
    \label{tab:col_head}
\end{table*}

\subsection{Combination into overall source classification}
\label{subsect:combination}
The above diagnostics provide two independent axes along which to infer the presence of an AGN. Firstly via an excess of radio emission beyond that expected for its SFR -- a method sensitive to all radio-loud AGN and some radio-quiet AGN via low-level jets or accretion disks (this is the primary method used to detect radiatively inefficient i.e. jet-mode AGN). Secondly, for those sources with the four diagnostic BPT emission lines measured in the Portsmouth catalogue, the presence of a radiative AGN source becomes apparent via its emission line ratios. The combination of these two axes then allows us to classify every emission-line source in our sample into one of the following four physical categories: 1) SFGs, 2) RQAGN (containing both Seyferts and LINERS), and two classes of radio-excess AGN - 3) high-excitation radio galaxies (HERGs), and 4) emission-line low-excitation radio galaxies (LINELERGs)\footnote{Note that in order to be placed on the BPT diagram sources must have measured emission lines, and so we refer to this class as ``LINELERGs" - likely to be a subset of the full LERG population (see Sections \ref{subsect:cat_desc} and \ref{sec:demographics})}. Table \ref{tab:col_head} shows how the results of these two diagnostics are combined into each of the physical classes listed above. As discussed in Section \ref{subsect:cat_desc}, we release with this paper our full table of diagnostics, allowing the user to construct a particular physical class of objects according to their own criteria. For instance the population of LERGs can be considered in a number of ways - while some LERGs are found to be quiescent galaxies with large jets, others are seen to exhibit emission lines consistent with being LINERs, and recent work in the LoTSS Deep Fields has identified a subset of star-forming LERGs \citep{Kondapally2022}.

\begin{figure*}
\includegraphics[width=\columnwidth]{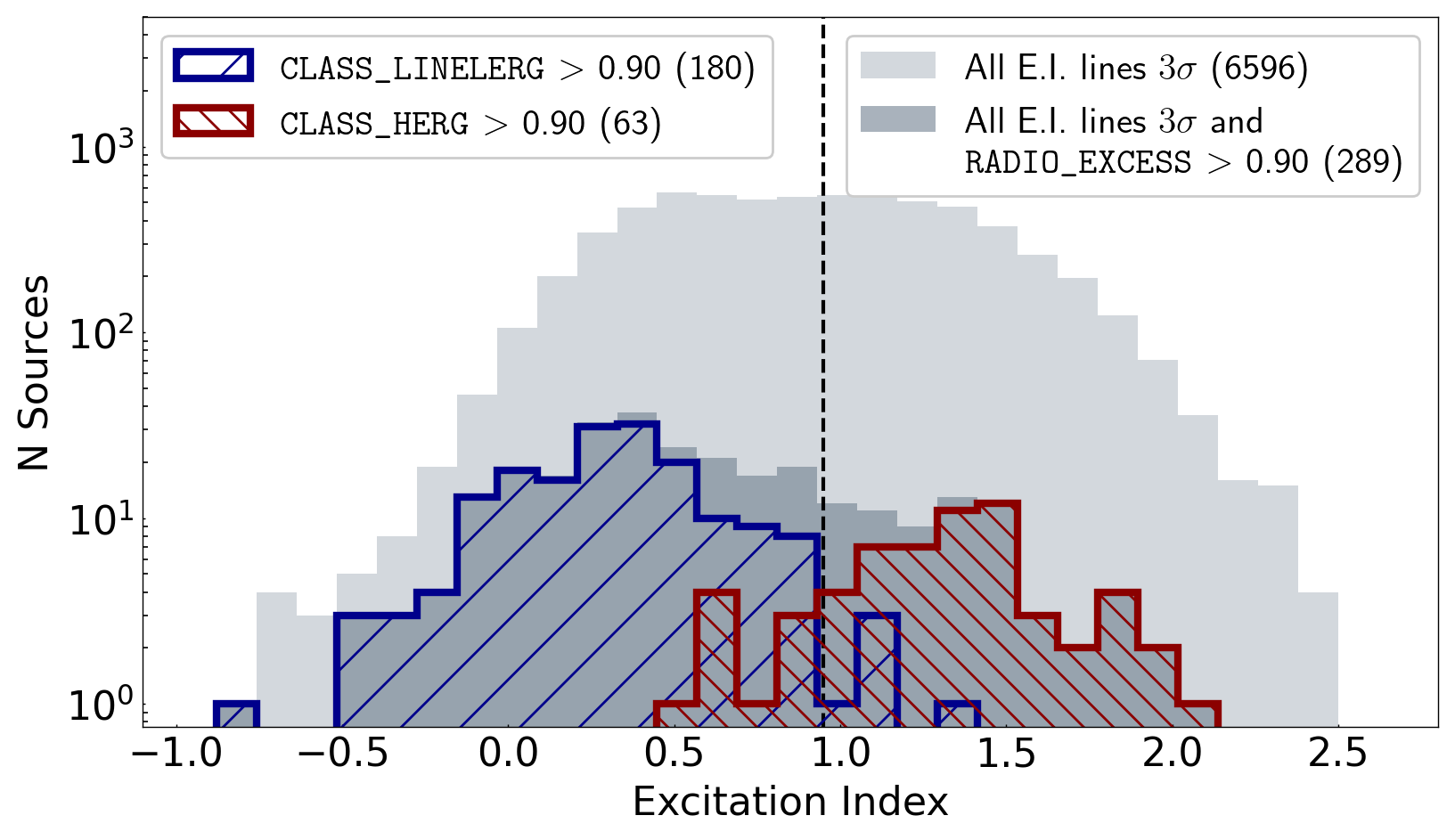}
\includegraphics[width=\columnwidth]{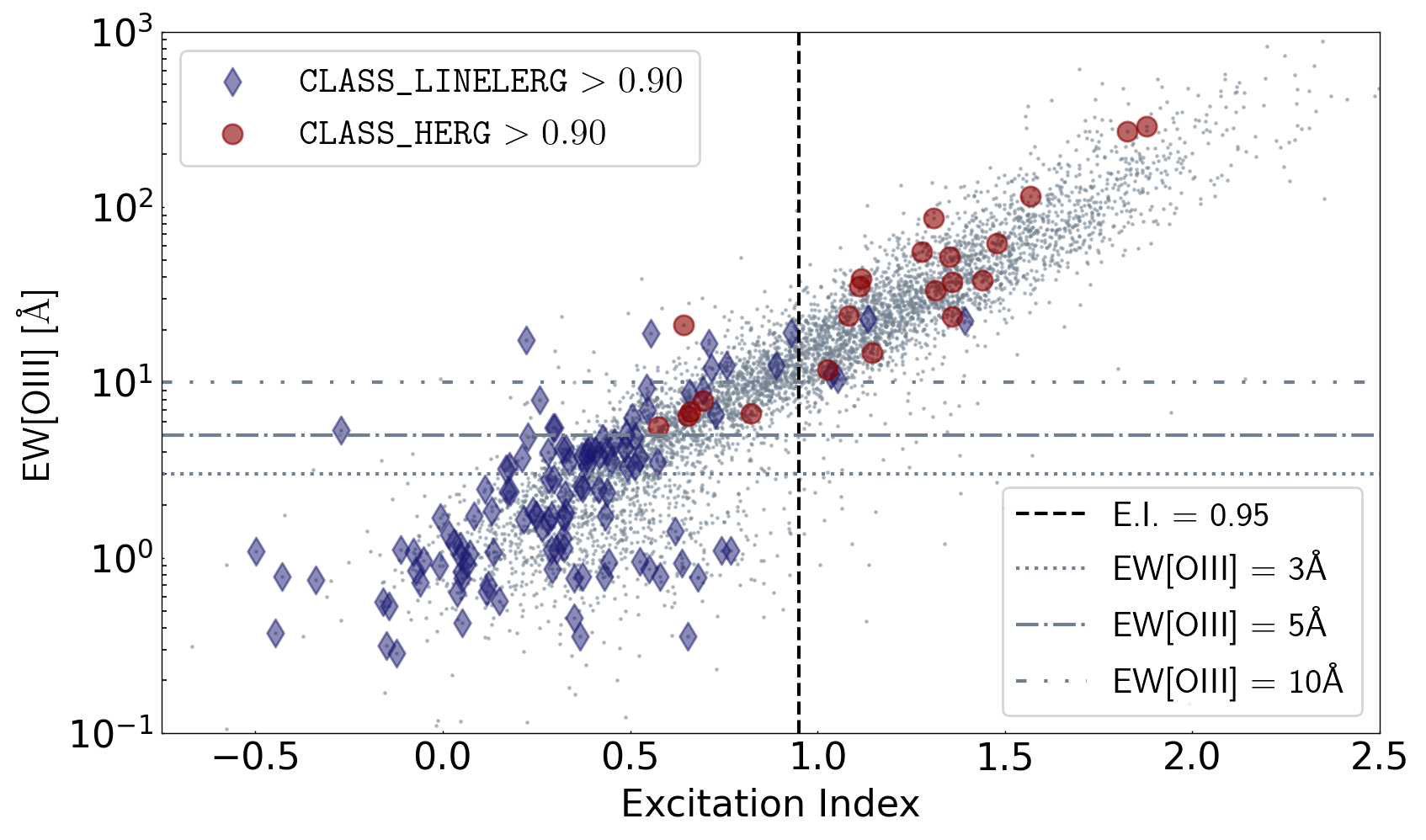}
    \caption{Left: The distribution in Excitation Index for all sources with $\ge3\sigma$ detections in all of the necessary emission lines (see text; shown in light grey) overlaid with the subsets of sources which have \texttt{RADIO\_EXCESS} $> 0.90$ (dark grey), \texttt{CLASS\_LINELERG} or \texttt{CLASS\_HERG} $> 0.90$ (shown as blue and red hatched histograms, respectively). The vertical dashed line at E.I.$=0.95$ indicates the threshold used by \citet{Buttiglione2010} to distinguish between high- and low-excitation sources. The legend indicates the number of sources in each class. Right: the distribution of sources in the E.I.-EW[O{\sc{iii}}] plane. In grey we show objects in our catalogue with a match in the MPA-JHU catalogue \citep{Brinchmann2004}, again with all E.I. emission lines at $>3\sigma$. We overlay the positions of the subsets of these objects with \texttt{CLASS\_LINELERG} and \texttt{CLASS\_HERG} showing a reliability $>0.90$. Horizontal dashed lines denote the various suggested divisions in the literature ~\protect\cite{Laing1994} at $3$\AA, ~\protect\cite{Best2012} at $5$\AA, and ~\protect\cite{Tadhunter1998} at $10$\AA. }
    \label{fig:ei}
\end{figure*}

\begin{figure*}
   \includegraphics[width=\textwidth]{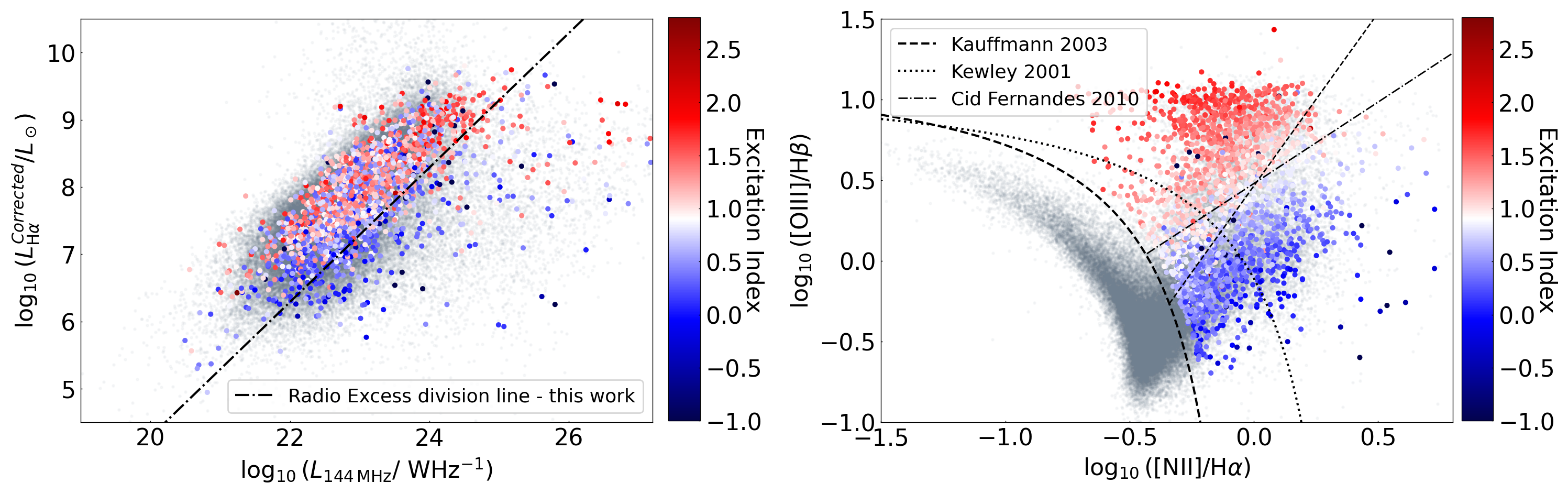}
    \caption{The location of sources in the diagnostic diagrams used for our classification (i.e. the H$\alpha$ vs 144\,MHz luminosity diagram and the BPT-[\textsc{Nii}] shown as a function of excitation index, indicated by the colour bar to the right for sources with \texttt{BPT\_SFG} $<0.1$. The distribution for the full population is shown for reference in the background in light grey to give context. In the right panel, the diagnostic lines from \citet{Kauffmann2003}, \citet{Kewley2001} and \citet{Cid2010} are shown as the dashed, dotted, and dot-dashed lines, respectively. The dot-dashed line in the left-hand panel represents the dividing line above which we identify a source as having a radio excess, given in equation \ref{eq:excess}.}
    \label{fig:dd_ei}
\end{figure*}

\subsection{Probabilistic classification of the faint radio source population}
\label{subsect:probabilistic_classif}
An important aspect of this work is our treatment of the uncertainty associated with the classification of objects into particular classes. As all measurements used in our analysis are accompanied by uncertainties (i.e. flux errors on the radio flux at 144 MHz and on all measured emission lines in the Portsmouth catalogue) we can simply perform a Monte-Carlo simulation to evaluate the range of possible outcomes. We generate 1000 realisations of the full catalogue, by perturbing the radio and line fluxes in each realisation by randomly drawing samples from normal distributions with standard deviation equal to the reported uncertainties. We then apply the full classification workflow to each of the thousand catalogues individually, before combining the results to determine what fraction of realisations of each source belong to each particular class. In addition to the overall source classifications we report the probabilistic results of each individual diagnostic - radio excess for all objects with a measured H$\alpha$ flux, and BPT line ratios for all objects with all four BPT diagnostic lines measured. In Figure \ref{fig:probs_by_class} we show the reliability distribution of the four physical classes following the construction of our probabilistic source classifications. The reliability reported along the horizontal axis depicts the fraction of realisations that have fallen into this class, such that each object is represented once in each of the four different classes (i.e. four times in total in this plot). An object which for instance is classified as an SFG in 80\% of the catalogue realisations will fall into the bin at a reliability of 0.8 in the grey histogram, if for instance in 10\% of realisations it is instead classified as a RQ AGN it will appear at a value of 0.1 in the gold histogram, and for example twice more at 0.05 in the blue and red histograms if the remaining 10\% of realisations results were distributed evenly between the HERG and LERG classes. One takeaway from Figure \ref{fig:probs_by_class} can be seen on the right-hand side of the plot; $\sim 40\,000$ sources are securely classified (in $>$90\% of realisations) as SFGs, $\sim 20\,000$ as RQ AGN, $\sim 7000$ as LINELERGs, and $\sim 400$ as HERGs\footnote{It is also interesting to note how the LINELERGs are divided between those with SFG-like and LINER-like emission. For objects with \classlinelerg\ $>0.90$, we find that for 50\% of realisations or more:  $\approx30$\% are classed as SFGs on the BPT diagram, $\approx64$\% are classed as LINERs, and the remainder switch between the SFG and LINER regions.}.

\subsection{Catalogue Description}
\label{subsect:cat_desc}
With this paper we release our probabilistic spectral source classifications for the \Nsrcs\ sources in common between LoTSS DR2 and SDSS-DR8/BOSS-DR12 with a measured H$\alpha$ flux. In Table \ref{tab:col_head} we list the column headings of our catalogue with a short description of each in turn. We provide three different warnings to caution the user of the potential pitfalls associated with a given object's classification. \texttt{CLASS\_z\_WARNING} provides an alert that the spectroscopic redshift recorded by Portsmouth differs from the \texttt{z\_best} estimate of \citealt{Hardcastle2023} (used to derive radio luminosities) by $\Delta z > 0.001$. The \texttt{BALMER\_CORR\_WARNING} indicates that our Balmer-correction on the H$\alpha$ luminosity has been carried out statistically as a consequence of a missing H$\beta$ flux in the Portsmouth catalogue (see Section \ref{subsect:rad_excess}). Finally we provide the \texttt{zscore} warning, to indicate the significance of deviation from the null hypothesis that maximum likelihood classification is correct. This warning applies to some low S/N spectra for which the nature of BPT classification (i.e. using ratios of line intensities) leads to non-intuitive behaviour in the Monte-Carlo realisations of the catalogue. For instance, when the flux measurement of one diagnostic line in a ratio is close to zero, but the other line is not, the realisations of this ratio may lie far from the measured line ratios (and consequently the maximum-likelihood classification). We provide in the \texttt{RADIO\_EXCESS} column a number between 0 and 1 indicating the number of catalogue realisations for which a radio excess was measured for each source in our catalogue. For those sources with all four of the diagnostic BPT lines recorded by Portsmouth we also provide a number between 0 and 1 in the \texttt{BPT\_SFG}, \texttt{BPT\_SEY}, \texttt{BPT\_LIN}, \texttt{BPT\_COMP}, \texttt{BPT\_CSEY}, and \texttt{BPT\_CLIN} columns indicating the fraction of times that each source was classified into these parts of the BPT diagram. We also include \texttt{BPT\_ML} which records the maximum-likelihood BPT classification of each source (i.e. where the Portsmouth-measured line ratios fall). Finally, in the \texttt{CLASS\_SFG}, \texttt{CLASS\_RQAGN}, \texttt{CLASS\_LINELERG} and \texttt{CLASS\_HERG} columns, we provide our probabilistic spectral source classifications into four physical classes. As we describe in Table \ref{tab:col_head}, the classes are constructed as follows in each individual realisation of the catalogue (before calculating their probabilistic ``average" results): \texttt{CLASS\_SFG} requires no radio excess to have been measured, and a BPT line ratio falling in the SFG part of the diagram. \texttt{CLASS\_RQAGN} requires that no radio excess is measured, and a BPT line ratio above the \cite{Kauffmann2003} demarcation line. \texttt{CLASS\_LINELERG} requires that a radio excess \emph{is} measured, and that the BPT line ratio lies in either the SFG or LINER regions. Finally to be classed as \texttt{CLASS\_HERG} requires that a radio excess \emph{is} measured, and that the BPT line ratio lies in the Seyfert region of the plot. \\

\noindent We also provide in the Appendix (Table \ref{tab:cat_example}) a snippet of the catalogue provided with this work, and some examples of how to construct a particular class of objects. \\

\begin{figure*}
    \includegraphics[width=\columnwidth]{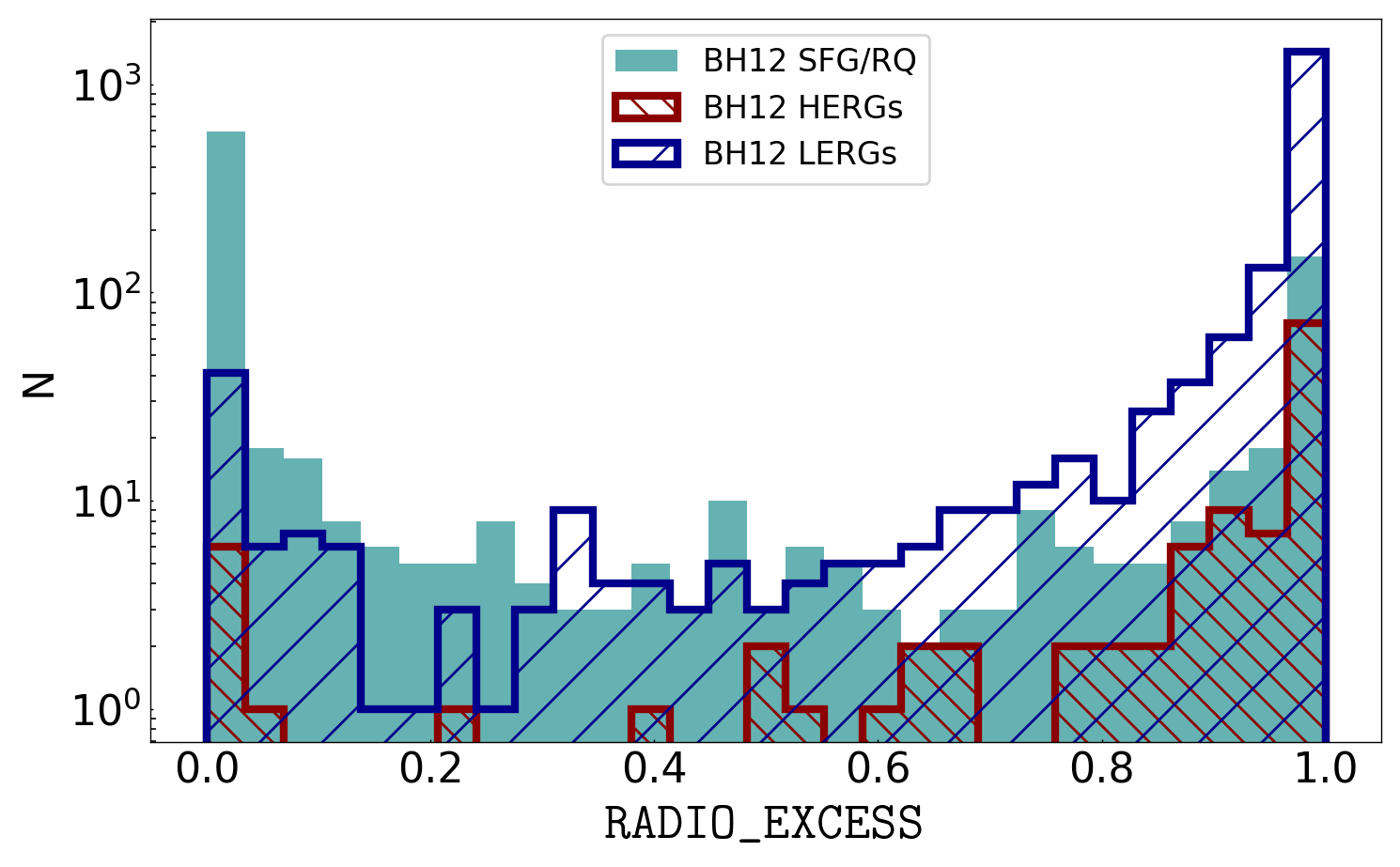}
    \includegraphics[width=\columnwidth]{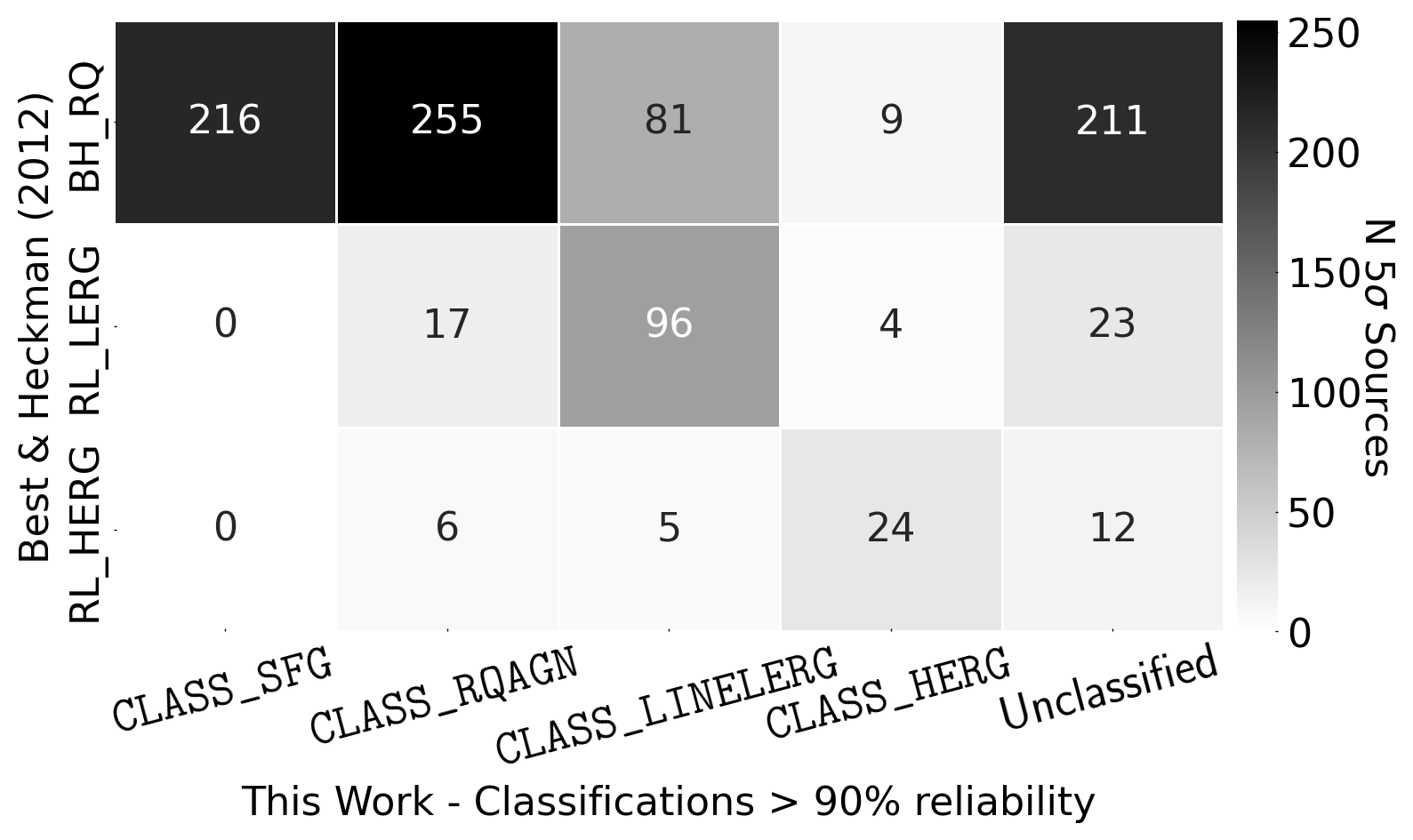}
    \caption{Left: The distribution of our \texttt{RADIO\_EXCESS} parameter for sources identified as SFG/radio-quiet objects, HERGs and LERGs in \citet{Best2012}, coloured teal, red and blue, respectively (as indicated in the legend). Right: a confusion matrix to compare final classifications of objects between BH12 and this work. The objects contained exhibit $5\sigma$ detections in all four BPT lines (in the Portsmouth catalogue), and have been classified with $>90$\% reliability in this work, with the remainder listed in the "unclassified" column. See text for details.}
        \label{fig:bnh_radio_excess2}
\end{figure*}

\section{Validation}
\label{Sect: Validation}

In this section, we will compare the results of our classifications with other works and diagnostics. 

\subsection{Excitation Index distribution for HERGs and LERGs}
\label{subsec:EI}

One of the key tools available for studying radio excess AGN is the Excitation Index \citep[hereafter EI;][]{Buttiglione2010} which is designed to measure the relative intensity of high- and low-excitation emission lines in a source's optical spectrum. Estimating a source's EI requires measurements of six emission line species (H$\beta$, \textsc{[Oiii]}$\lambda 5007$, \textsc{[Oi]}$\lambda 6364$, \textsc{[Nii]}$\lambda 6583$, H$\alpha$, and the \textsc{[Sii]}$\lambda 6716, 6731$ doublet), combined such that: 

\begin{equation}
    \begin{split}
        \mathrm{EI} \equiv \log_{10}(\textsc{[Oiii]}\slash \mathrm{H}\beta) -\frac{1}{3}[&\log_{10}(\textsc{[Nii]}\slash \mathrm{H}\alpha) + \log_{10}(\textsc{[Sii]}\slash \mathrm{H}\alpha) \\&+ 
        \log_{10}(\textsc{[Oi]}\slash \mathrm{H}\alpha)]
    \end{split}
    \label{eq:ei}
\end{equation}

\noindent with the dividing line between high- and low-excitation sources defined to be at EI = 0.95 (\citealt{Buttiglione2010}). We decided against adopting an EI criterion in our classification scheme since the requirement for detections of \textit{six} line species was likely to dramatically reduce the number of sources that we are able to classify. Nevertheless, since this is one of the key diagnostics of the ionisation state of radio excess sources \citep[e.g. in][]{Best2012} it is clearly of interest to see how our results compare for the subset of sources for which we have the necessary number of detections.

The distribution of Excitation Index for all $6\,596$ sources with the necessary six emission lines measured at $\ge 3\sigma$ is shown as the light grey histogram in figure \ref{fig:ei}, with the canonical dividing line between low- and high-excitation sources overlaid as the vertical dashed line. However, only $289$ of these sources are classified as having a radio excess (i.e. \texttt{RADIO\_EXCESS} $>\,0.90$ in our catalogue), and their EI distribution is overlaid in dark grey. This represents just $0.007$ percent of the $38\,588$ sources for which we are able to identify a radio excess at this reliability by comparing their H$\alpha$ and 144\,MHz luminosity. 

We have also overlaid the EI distributions of sources which our classification scheme identifies as having \texttt{CLASS\_HERG} and \texttt{CLASS\_LINELERG} $> 0.90$ as red and blue hatched histograms, respectively. It is encouraging to see that the mean EI for the HERGs and LINELERGs offset from one another, but also that the canonical dividing line of EI $= 0.95$ falls very close to the point where the two histograms cross. Indeed, using this subset with 3$\sigma$ detections in all six EI lines, we find that 84 percent of the HERGs have EI $> 0.95$, while 97 percent of the LINELERGs have EI $< 0.95$.

In the right-hand panel of Figure \ref{fig:ei} we show sources with matches in the MPA-JHU catalogue (\citealt{Brinchmann2004}) to demonstrate the distribution of EW[O{\sc{iii}}] values as a function of E.I. As previously, we show only those sources for which all six E.I. emission lines were measured with $>3\sigma$ significance, but this time they must also have a match in the MPAJHU catalogue (a total of $5\,931$ sources). Overlaying the subsets of these sources which are also identified as \texttt{CLASS\_HERG} or \texttt{CLASS\_LINELERG} $> 0.90$, we find that the sources scatter in approximately the expected locations. We have denoted on the plot some previously suggested divisions between the HERG and LERG populations in terms of [O{\sc{iii}}] equivalent width -- $3$\AA\ \citep{Laing1994} and $10$\AA\ \citep{Tadhunter1998}. The lower division line does seem to cleanly select securely-identified LERG-like objects, while the upper suggested division is more akin to a clean selection of HERG-like objects. Substantial mixing between securely-classified \texttt{CLASS\_HERG} and \texttt{CLASS\_LINELERG} objects is seen between the two lines. 

Finally, in Figure \ref{fig:dd_ei} we show the two diagnostics that we have used to classify our sample (the BPT diagram is shown in the left panel, with the radio excess plot shown in the right panel), where we have colour-coded all sources that are unlikely to be SFGs (i.e. those that have \texttt{BPT\_SFG} $< 0.1$) but with $\ge 3\sigma$ detections in the EI emission lines by their EI. The colour scaling has been chosen such that sources with EI $>0.95$ are shown in red, and those with EI $< 0.95$ are in blue. It is immediately apparent that the areas of the BPT diagram that we call `high-excitation' (namely the Seyfert area) are dominated by sources with EI$>0.95$, and conversely that the LINER area (lower right in the BPT diagram, which we define as containing low-excitation sources) are dominated by sources with blue colours, signifying $EI < 0.95$. This was also noted by \citet{Buttiglione2010}, however, the sample used in that work contained only 113 bright sources \citep[building on the previous works of][]{Hine1979,Laing1994}, which meant that it was not possible to compare the two methods at the time.

Finally we note that there is an interesting possible correlation between the position in the H$\alpha$ vs 144\,MHz luminosity plot and the EI for the sources that do not show a radio excess. It appears as though the sources with lower H$\alpha$ luminosity (and therefore lower SFRs assuming a linear correlation, e.g. \citealt{kennicutt2012}) appear to have weaker high-excitation lines than the higher H$\alpha$ luminosity (more vigorously star-forming) counterparts. This is in keeping with expectations that star formation and radiative AGN activity go hand-in-hand, as noted by e.g. \cite[][]{gurkan2015}.

This good agreement between EI and our method offers significant encouragement that we are able to identify high-excitation sources, even in the absence of detections in all six lines required for calculating the EI.

\begin{figure*}
    \includegraphics[width=0.99\textwidth]{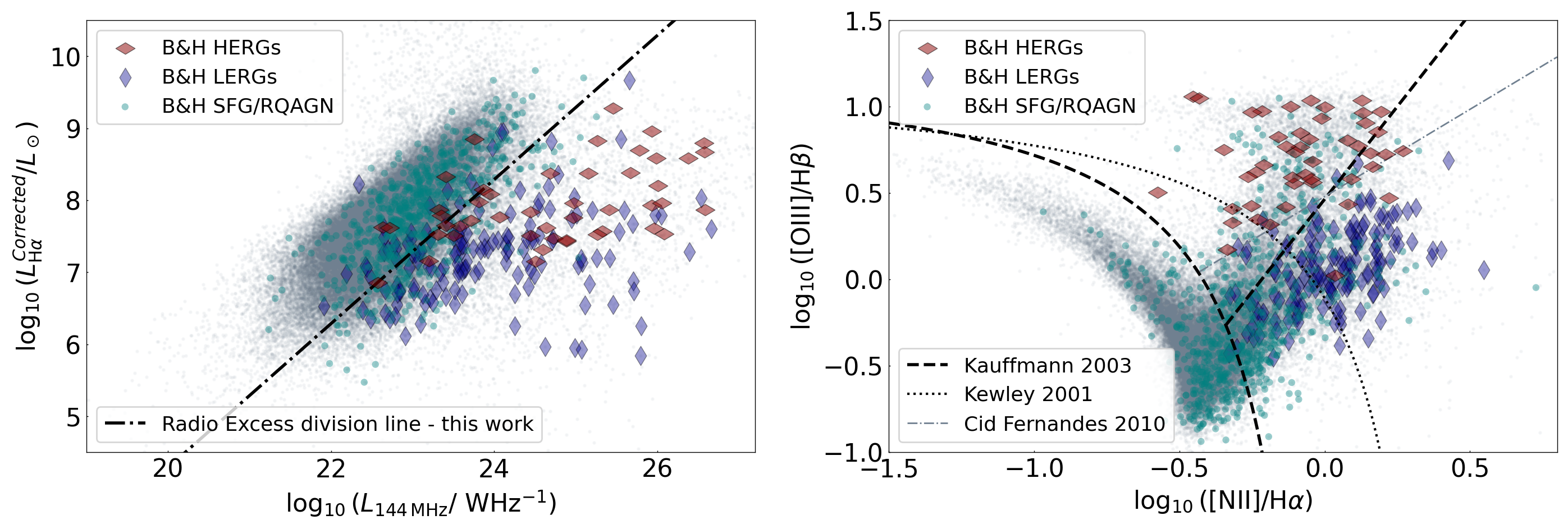}
    \caption{The L$_{{\rm{H\alpha}}}$ vs L$_{{\rm{144}}}$ (left), and BPT (right) parameter spaces for sources with $5\sigma$ detections in all four BPT lines in the Portsmouth catalogues. We overlay sources identified as SFGs/radio-quiet, HERGs and LERGs in \citet{Best2012}, coloured teal, red and blue, respectively (as indicated in the legend). The full population of 5$\sigma$ sources in our catalogue is shown as the grey distribution in the background.}
    \label{fig:bnh_why}
\end{figure*}

\begin{figure*}
    \includegraphics[width=\columnwidth]{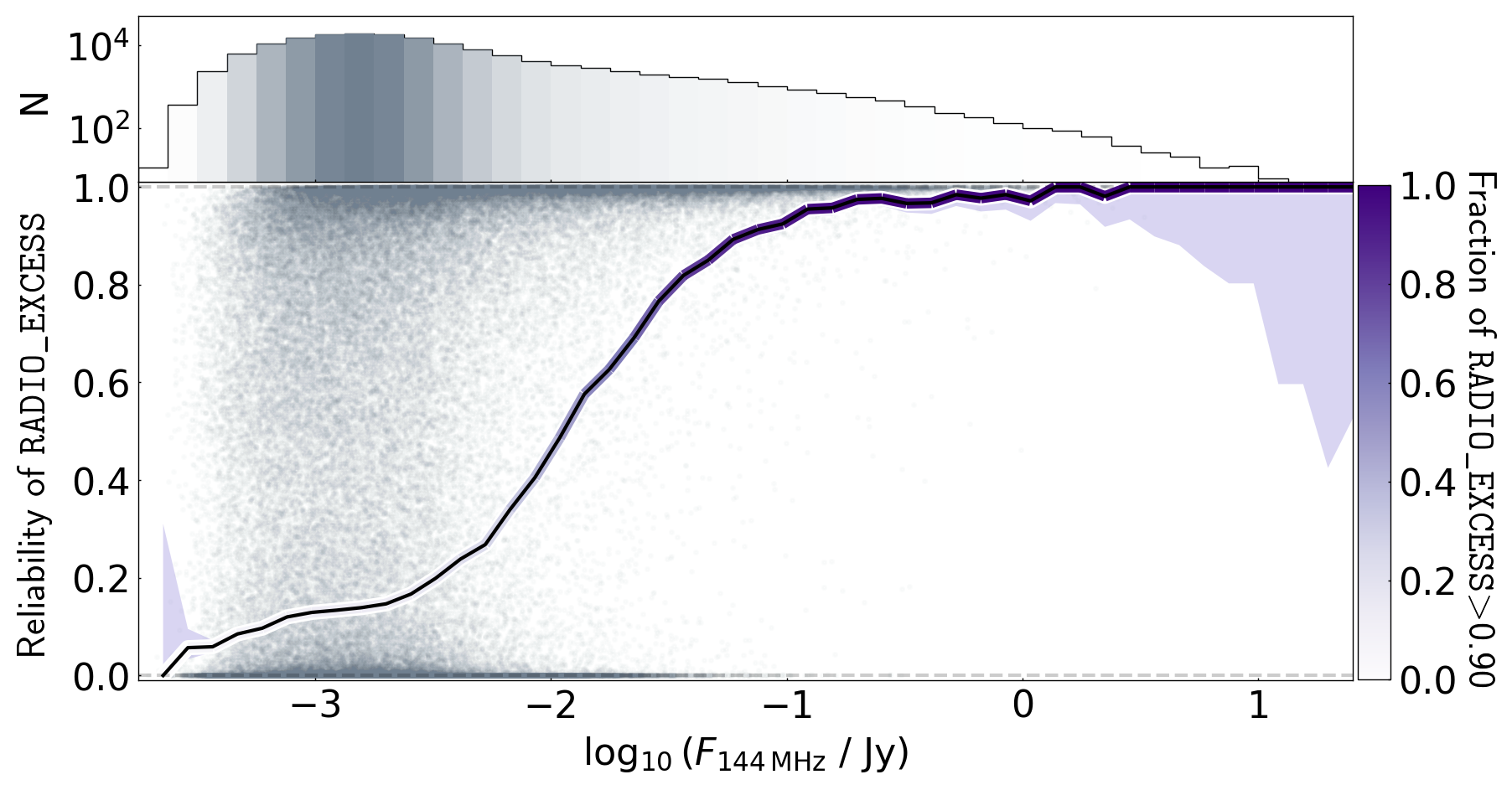}
    \includegraphics[width=\columnwidth]{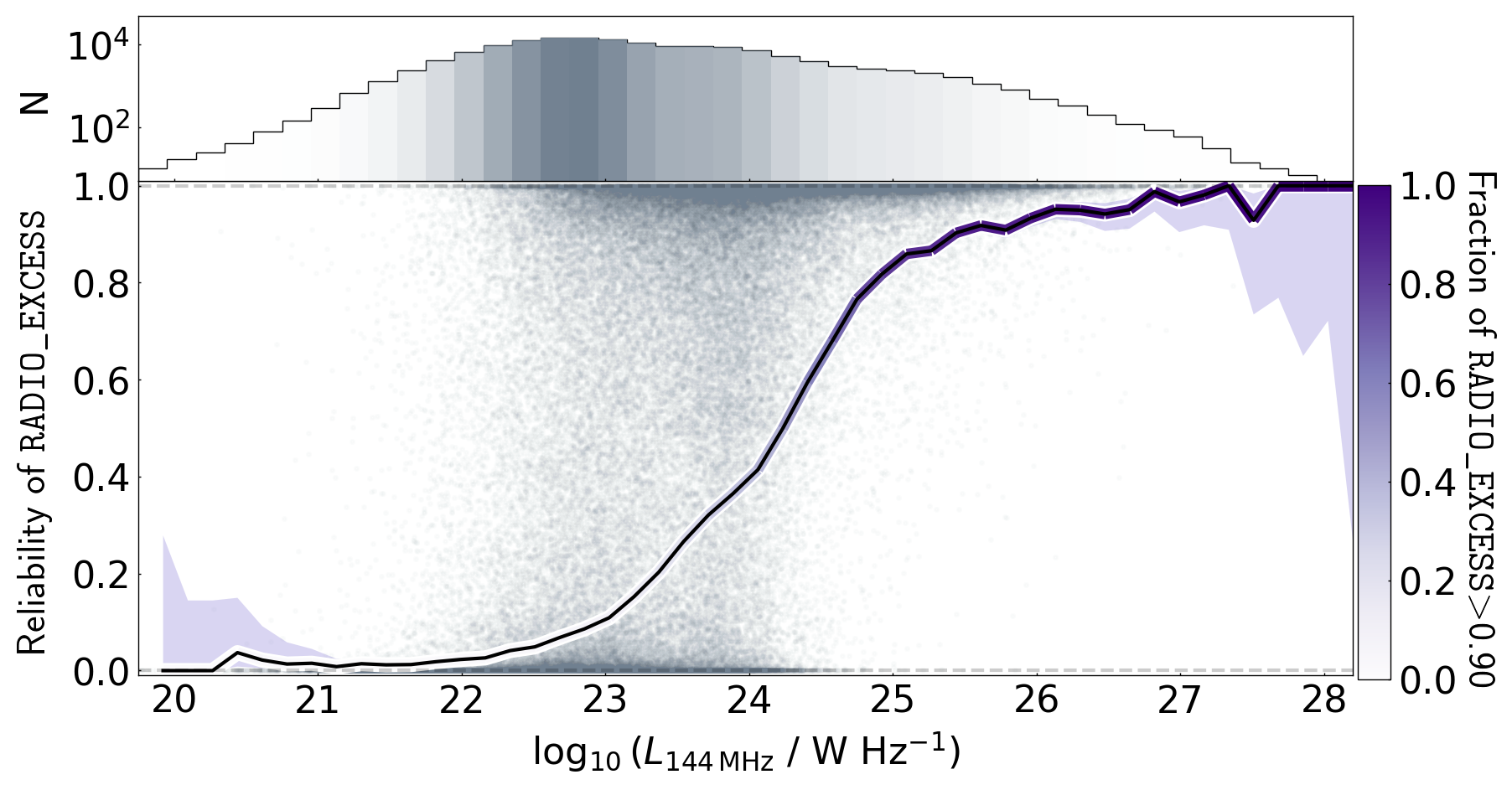}
    \caption{The variation in the \texttt{RADIO\_EXCESS} parameter as a function of (left:) 144\,MHz flux density and (right:) luminosity. In the main panels the greyscale scatter points indicate the source poistions in this plane, and these have been overlaid with the cumulative count of sources passing the 90\% reliability threshold in our \radioexcess\ parameter in bins of flux density\slash log luminosity. The purple shading indicates the 90 per cent confidence interval on this fraction. The upper panels indicate the overall distributions of flux density and luminosity, on the same horizontal axes. The histograms are shaded to scale linearly with the number of sources in each bin.}
    \label{fig:radio_excess_by_flux_and_lum}
\end{figure*}

\subsection{Comparison with Best \& Heckman catalogue}
\label{subsec:bnh2012}

An additional test of how well our method performs, is to compare our results with \cite[][hereafter BH12]{Best2012}. To do this, we first identify common sources by performing a positional cross match between the BH12 source positions, and the positions of optical counterparts to the LoTSS sources in the \cite{Hardcastle2023} catalogue, using a nearest neighbour algorithm with maximum search radius equal to 1 arcsec. We find that $4\,772$ sources satisfy the matching criteria ($3\,279$ of with all four BPT lines recorded in the Portsmouth catalogue enabling a full classification in this work). This includes 116 BH12 HERGs, 1864 LERGs and 927 sources which BH12 classified as `star forming galaxy' (however we note that in B12 this classification also includes radio-quiet AGN, on the basis that they assume a stellar origin for the radio emission in those sources). 

The distribution of our \texttt{RADIO\_EXCESS} parameter for sources classified as SFG, HERG and LERG by BH12 are shown as the solid teal, blue and red hatched histograms in Figure \ref{fig:bnh_radio_excess2}. It is very encouraging to note that $86$ percent of the sources that B12 classify as RLAGN are assigned \texttt{RADIO\_EXCESS} values $> 0.90$ in our catalogue.

However when we look at how the different BH12 source types are classified overall (i.e. including the BPT information), it doesn't look quite as encouraging: 

\begin{itemize}
\item 55\% of the 927 BH12 SFGs/Radio-quiet objects have \texttt{CLASS\_SFG} or \texttt{CLASS\_RQAGN} $> 0.90$,

\item 25\%  of the 116 BH12 HERGs have \texttt{CLASS\_HERG} $> 0.90$, and

\item 58\% of the 1864 BH12 LERGs have \texttt{CLASS\_LINELERG} $> 0.90$.
\end{itemize}

The BH12 classification scheme includes a complex workflow that first attempts to classify sources as high- or low-excitation on the basis of their EI, BPT line ratios (using the standard  [N{{\sc{ii}}}]/H$\alpha$ and [O{\sc{iii}}]/H$\beta$ axes) and EW$_\mathrm{[OIII]}$ properties being at least $1\sigma$ away from some criterion. If a source does not meet the 1$\sigma$ criterion in all of these three categories, then the steps are repeated without the 1$\sigma$ criterion. This is a pragmatic approach to classifying as many sources as possible, but also means that the BH12 classifications are maximum-likelihood estimates, and often have significant uncertainties associated with them that are not visible \textit{a posteriori}, for instance when using the catalogue. To provide a fairer comparison of the two approaches, we repeat the analysis above, this time considering only those sources which have $5\sigma$ detections in each of the four BPT lines in the Portsmouth catalogue in order to identify those sources which have the most reliable classifications in the BH12 catalogue. If we do this, we find that: 

\begin{figure*}
    \includegraphics[width=\textwidth]{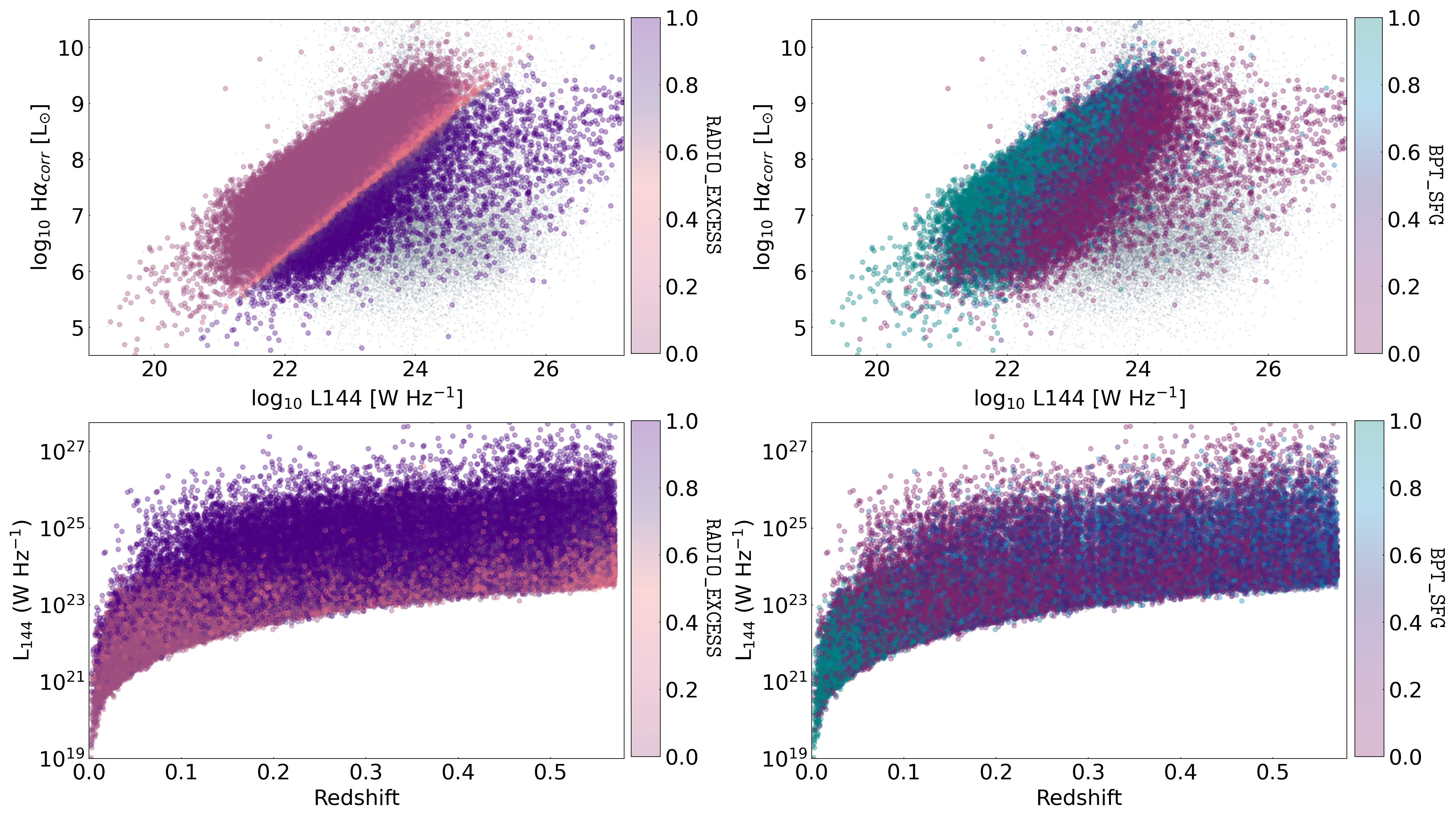}
    \caption{Top left (right) The 144\,MHz--H$_\alpha^\mathrm{corrected}$ Luminosity plane colour-coded by the 
    \texttt{RADIO\_EXCESS} (\texttt{BPT\_SFG}) parameter, i.e. the fraction of realisations for which we identify a source as having a radio excess (/being identified as a radiative AGN on the BPT diagram). These two quantities represent the two independent axes along which we identify signatures of AGN activity. In the lower left (right) panel we show the relationship between $z$ and L$_\mathrm{144\,MHz}$,
    again colour coded by \radioexcess\ (\texttt{BPT\_SFG}).}
    \label{fig:z_vs_L_classified}
\end{figure*}

\begin{itemize}
    \item 61\% (67\%) of the 772 BH12 SFG have \texttt{CLASS\_SFG} or \texttt{CLASS\_RQAGN} $> 0.90$ ($> 0.80$), 
    \item 51\% (60\%) of the 47 BH12 HERGs have \texttt{CLASS\_HERG} $> 0.90$ ($> 0.80$), and
    \item 69\% (71\%) of the 140 BH12 LERGs have \texttt{CLASS\_LINELERG} $> 0.90$ ($> 0.80$).
\end{itemize}

Taking this $5 \sigma$ sample, and considering only those sources which are "reliably" classified in our catalogue (reliability $>90\%$) we construct the confusion matrix shown in the right-hand panel of Figure \ref{fig:bnh_radio_excess2}. This gives us chance to examine the classification of sources for which we would like to make a reliable (and correct) classification. Considering the top row first, BH12 $5\sigma$ ``SFGS"/radio-quiet objects (772 sources), it is encouraging to see that 216 and 255 of these objects are encapsulated by our \texttt{CLASS\_SFG} or \texttt{CLASS\_RQAGN} classes respectively. A further 90 objects have been reliability classified into one of the radio-excess classes (81 in \texttt{CLASS\_LINELERG} and 9 in \texttt{CLASS\_HERG}) - as we have deliberately re-defined the threshold above which we consider objects to have a radio-excess (with a physically-motivated division based on the distribution of L$_{{\rm{144}}}$-L$_{\rm{H\alpha}}$ values) this is not surprising. The remainder of the 772 BH12 $5\sigma$ ``SFGS"/radio-quiet objects (211 sources) evidently have reliabilities in each class \textless 90\% reflecting the uncertainty in their classification. This is the expected behaviour from our approach and quantifies the uncertainty on classifications that might be considered secure via S/N alone. In the central row, of BH12 $5\sigma$ LERGs (140 sources), it is reassuring to see that 96 objects have been reliably classified into \texttt{CLASS\_LINELERG}. 17 objects however have shifted to being reliably classified as \texttt{CLASS\_RQAGN}, which, similarly to above, can be attributed to our updated definition of the division between radio-quiet and radio-excess objects. No objects have been classified reliably as \texttt{CLASS\_SFG}. Perhaps more puzzlingly, 4 objects are reliably classified into \texttt{CLASS\_HERG} in this work. These objects are not ambiguously placed on the BPT diagram, nor do they exhibit an E.I. which would class them as a HERG in BH12. It is true however that BH12 were using a previous version of an SDSS catalogue, and so small differences in flux measurements (or their uncertainties) may shift these objects across diagnostic boundaries or into subsequent parts of the BH12 workflow. Again the remainder of the 140 $5\sigma$ BH12 LERGs (23 sources) have been classified with reliabilities \textless 90\% in any class, reflecting the ambiguity in their ultimate classification. In the bottom row of BH12 $5\sigma$ HERGs (47 sources), 24 sources are reliabily classified as \texttt{CLASS\_HERG}, another 6 objects have shifted to \texttt{CLASS\_RQAGN} due to the updated radio-excess threshold, and 5 objects have been reliably classified as \texttt{CLASS\_LINELERG}. The shifts between \texttt{CLASS\_HERG} and \texttt{CLASS\_LINELERG} are not intuitive, and are most likely to have arisen from differences in the reported flux measurements of the two different catalogues. \\

Figure \ref{fig:bnh_why} shows the location of this BH12 5$\sigma$ sample in our diagnostic diagrams. In both plots, the full population in our catalogue with $\ge 5\sigma$ detections in all four BPT emission lines is shown in light grey, and has been overlaid with the BH12 “SFGS"/radio-quiet objects, LERGs and HERGs (shown as teal, blue and red symbols, respectively). This is useful information since, as touched-upon above, in addition to the different workflow implemented, BH12 used the MPA-JHU catalogue from \citet{Brinchmann2004} rather than the later updated Portsmouth catalogue described by \citet{Thomas2013}. It is certainly possible for line measurements to differ between the two catalogues given the significantly different methodologies, however, we see excellent general agreement, with BH12 HERGs and LERGs residing predominantly in the expected regions of the BPT diagram, and the BPT SFGs consistent with a mix of sources which are genuinely SFG (according to our definition) and those which host a radiative AGN but no radio excess. 

Comparing the H$\alpha$ and radio luminosity methods between the two works is more complicated, since (aside from using a different radio frequency of 1.4\,GHz from NVSS) BH12 also compare the strength of the 4000\AA\ break (measured using the $D_n(4000)$ parameter in the MPAJHU catalogue) with the stellar-mass-normalised 1.4\,GHz luminosity \citep[see also][]{Sabater2019}. Nevertheless, it is clear that the great majority of BH12 HERGs and LERGs lie in the radio-excess area of this diagnostic diagram (to the right of figure \ref{fig:bnh_why}), while the majority of the BH12 SFGs reside in the region expected for SFGs.

\begin{figure*}
\centering
    \includegraphics[width=0.49\textwidth]{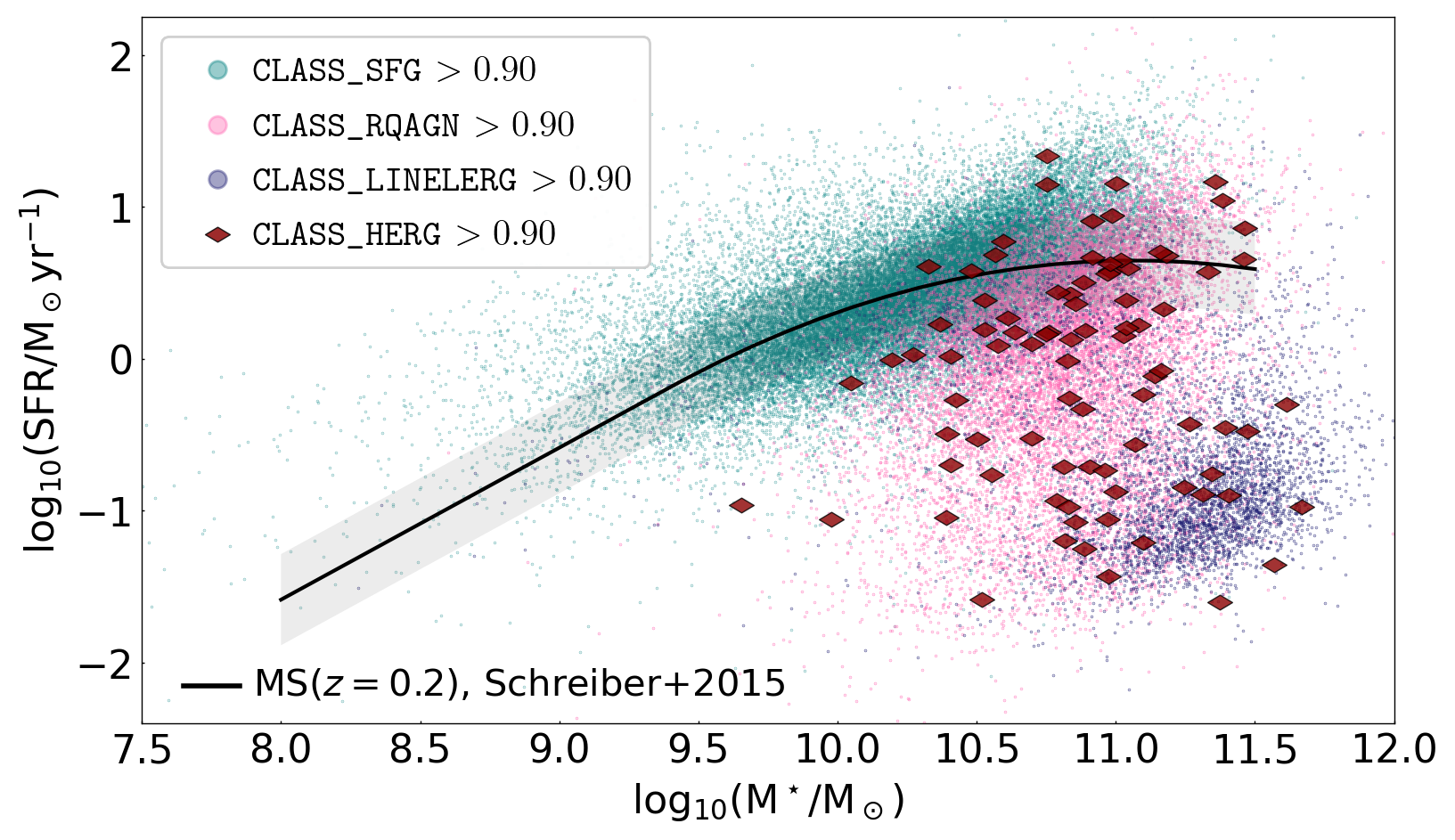}
    \includegraphics[width=0.49\textwidth]{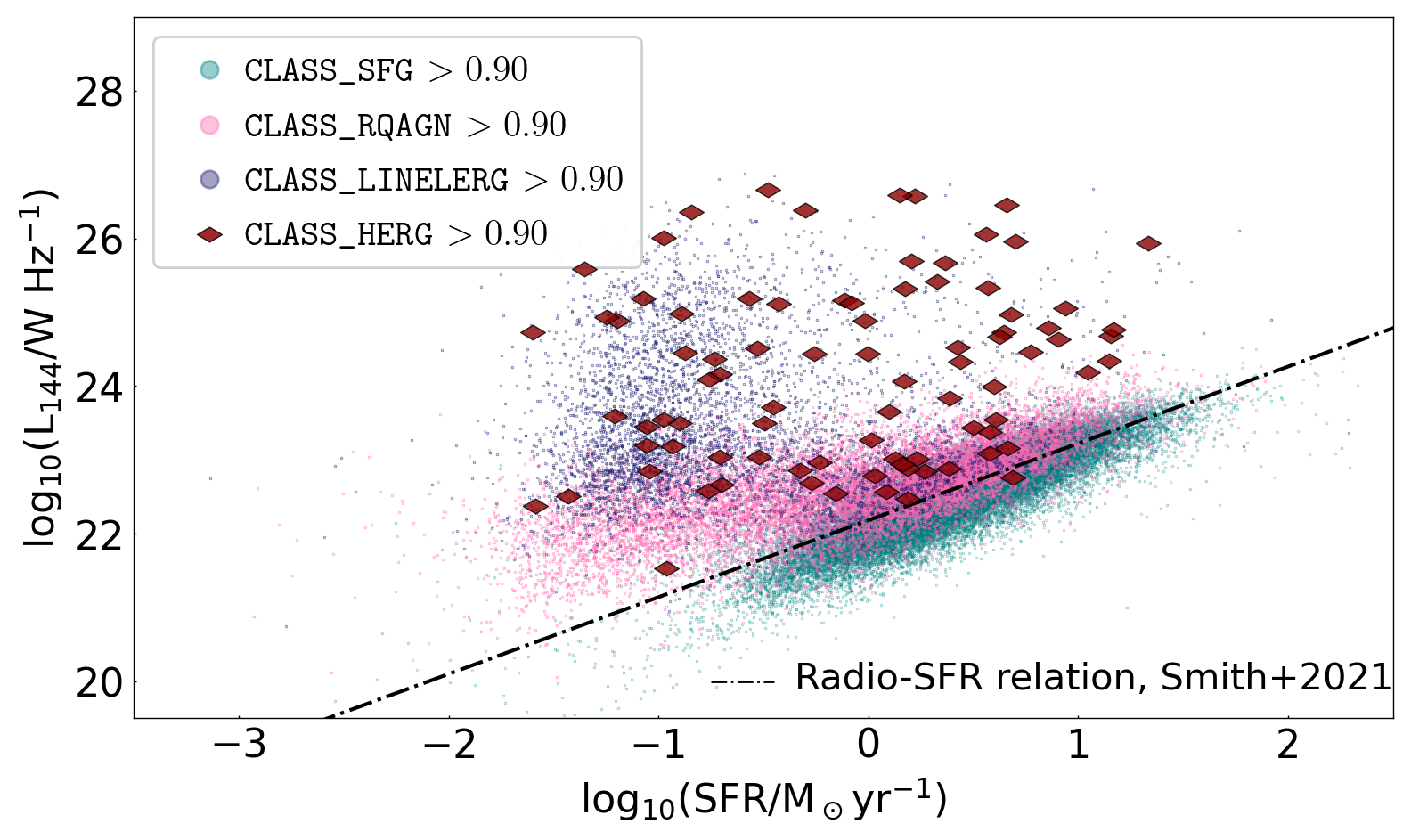}
    \caption{The distribution in the (Left:) SFR--stellar mass and (Right:) SFR--$L_\mathrm{144\,MHz}$ planes for the radio sources which have $>90$\,percent reliability in each class (SFG, RQAGN, HERG and LINELERG). SFGs are indicated by the teal points, RQAGN as pink, and LINELERGs in dark blue. HERGs are overlaid as larger dark-red diamonds. In the left panel, the ``main sequence'' relation from \citet{Schreiber2015} evaluated at $z = 0.2$ along with a 0.3\,dex scatter is indicated by the black line, for context. Similarly, in the right panel, the relationship between SFR and 144\,MHz luminosity from \citet{Smith2021} is shown as the lack dot-dashed line in the right panel. The SFR and stellar mass estimates come from the MPAJHU catalogue as an independent validation of our method.}
    \label{fig:sfgs_plus_contours}
\end{figure*}

\begin{figure}
\includegraphics[width=0.99\columnwidth]{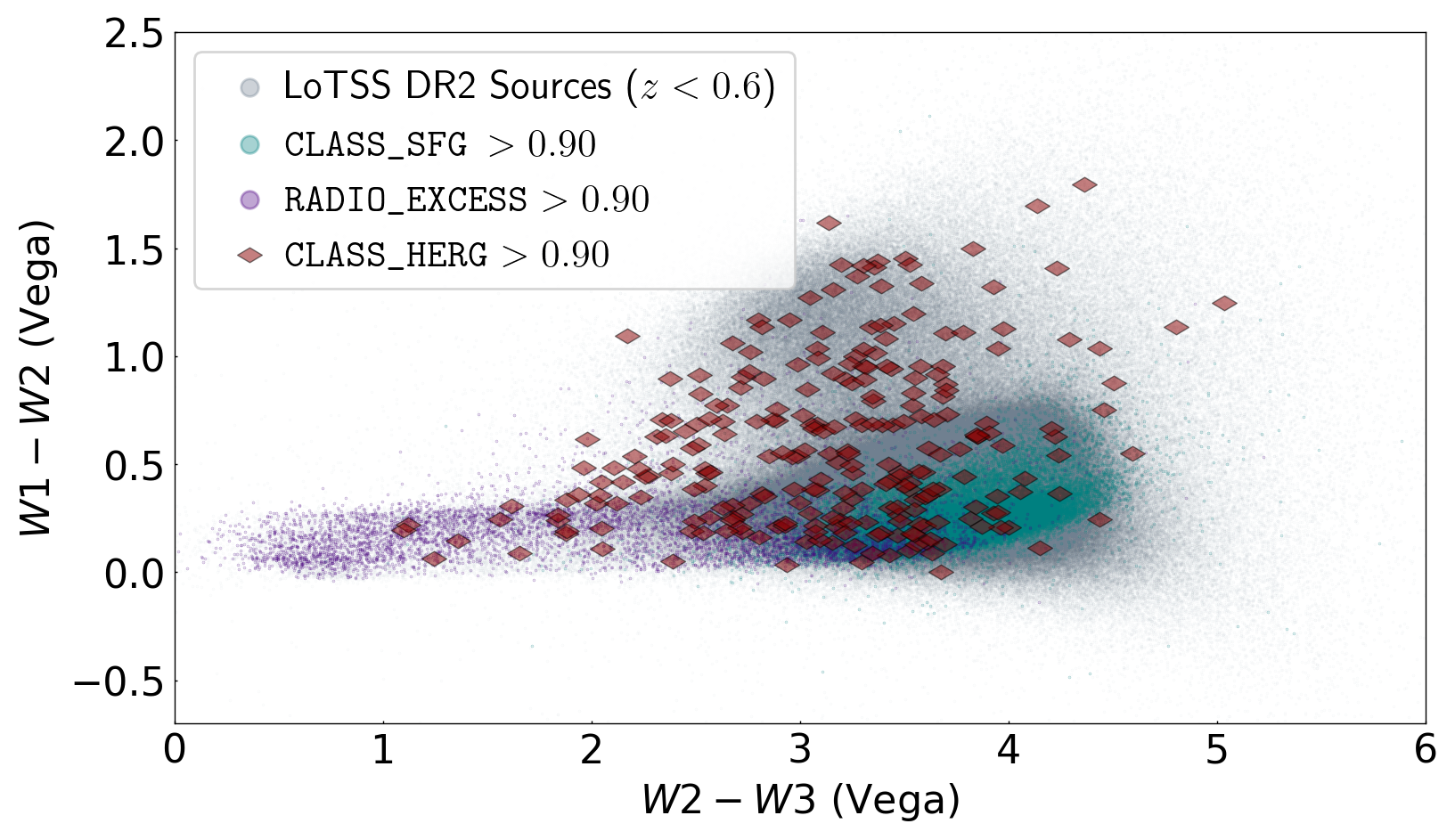}
\caption{{\textit WISE} colour-colour plot in Vega magnitudes for objects with good {\it WISE} photometry. The logarithmic density plot in grey shows 1.3 million DR2 objects with good photometry from \citet{Hardcastle2023}. Colours indicate high-confidence SFGs, radio-excess sources and the HERG subset of radio-excess objects.}
\label{fig:wisecc}
\end{figure}

\subsection{Further tests of probabilistic classification}
\label{subsect:further_tests}
Figure \ref{fig:radio_excess_by_flux_and_lum} shows the relationship between our \radioexcess\ parameter (i.e. the fraction of realisations for a given source that have a 144\,MHz luminosity excess) and the 144\,MHz flux density and luminosity, in the left and right panels respectively. The individual source properties are shown as the greyscale points in the background, and these have been overlaid with the cumulative count of sources passing the 90\% reliability threshold in our \radioexcess\ parameter, with a 90 percent confidence interval indicated by the purple shading. As expected, we find that the brightest and most luminous sources virtually all exhibit a radio excess, consistent with our expectations that the brightest sources should be predominantly RLAGN \citep[e.g.][]{Best2023,Whittam2022}. Similarly, the radio-excess fraction is much smaller among the faintest and least-luminous sources. In the upper panels we also include histograms to depict the distribution of sources in terms of their 144\,MHz flux density and log luminosity on the left and right respectively. The histogram's shading scales linearly with the number of sources - this is particularly of use in the right-hand panel, as it becomes evident that the majority of sources reside in the log L$_{\rm{144}} \approx 23$ region, which is not necessarily clear in the scatter plot.

In the top panels of Figure \ref{fig:z_vs_L_classified} we show the comparison between H$\alpha$ and 144\,MHz luminosity for all sources in our sample, which are shown as the grey background image. Overlaid are the sources with $\ge 5\sigma$ detections in both the 144\,MHz LoTSS data and the H$\alpha$ emission line, and these have been colour-coded by the \radioexcess\ and \texttt{BPT\_SFG} parameter in the left and right panels respectively. By design, we see a larger radio excess as we move to the right side of the left-hand side plot, however, unlike previous works, our Monte Carlo simulations have enabled us to determine the reliability of a radio excess for every source accounting for the uncertainties in both quantities. In the right-hand panel the distribution of \texttt{BPT\_SFG} is as we would expect; towards more reliable values of \texttt{BPT\_SFG} objects cluster in the radio-quiet portion of the plot, and at the lowest values of \texttt{BPT\_SFG} (i.e. selecting all radiative AGN) sources dominate the plot towards the highest radio luminosities, and are intermingled with the radio-quiet portion of the plot due to the radio-quiet radiative AGN selected via the BPT diagram. 

In the lower panels of Figure \ref{fig:z_vs_L_classified} we show how our sample fills the redshift vs 144\,MHz luminosity plane. Each point has again been colour-coded by the \radioexcess\ (left) and \texttt{BPT\_SFG} (right) parameter. As expected on the basis of Figure \ref{fig:radio_excess_by_flux_and_lum}, in the left-hand panel we see that radio excess sources are more prevalent at higher 144\,MHz luminosities across virtually the full redshift range and virtually absent at the lowest redshifts in our sample. An interesting feature is visible in the \radioexcess\ values visible for sources at the highest end of our redshift range - the reduced certainty over whether sources are radio excess could be a result of the lower SNR ratio in the H$\alpha$ fluxes that we have been able to measure for the most distant BOSS sources. In the right-hand panel we show the same redshift--luminosity plane; however, this time sources have been colour-coded by their \texttt{BPT\_SFG} values (i.e. the fraction of the 1000 Monte Carlo realisations of their emission line and 144\,MHz properties for which they are classified as star-forming galaxies on the standard BPT plot). As expected, the majority of the SFGs identified with the highest confidence (i.e. \texttt{BPT\_SFG} $> 0.9$) are local and it becomes increasingly difficult to identify SFGs at high confidence above $z \approx 0.25$ (see also section \ref{sec:demographics}, below).

\begin{figure*}
    \includegraphics[width=0.48\textwidth]{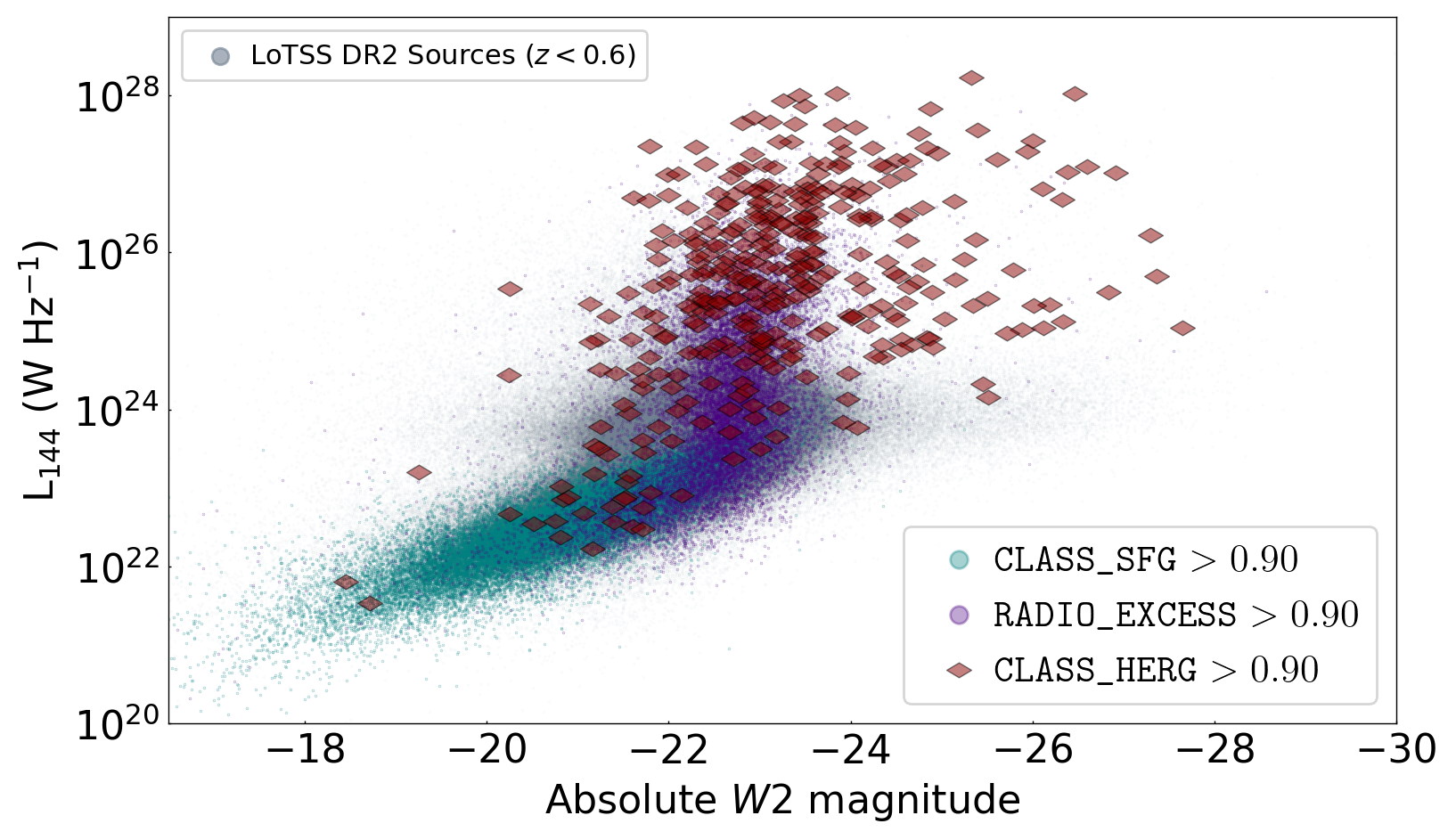}
    \includegraphics[width=0.48\textwidth]{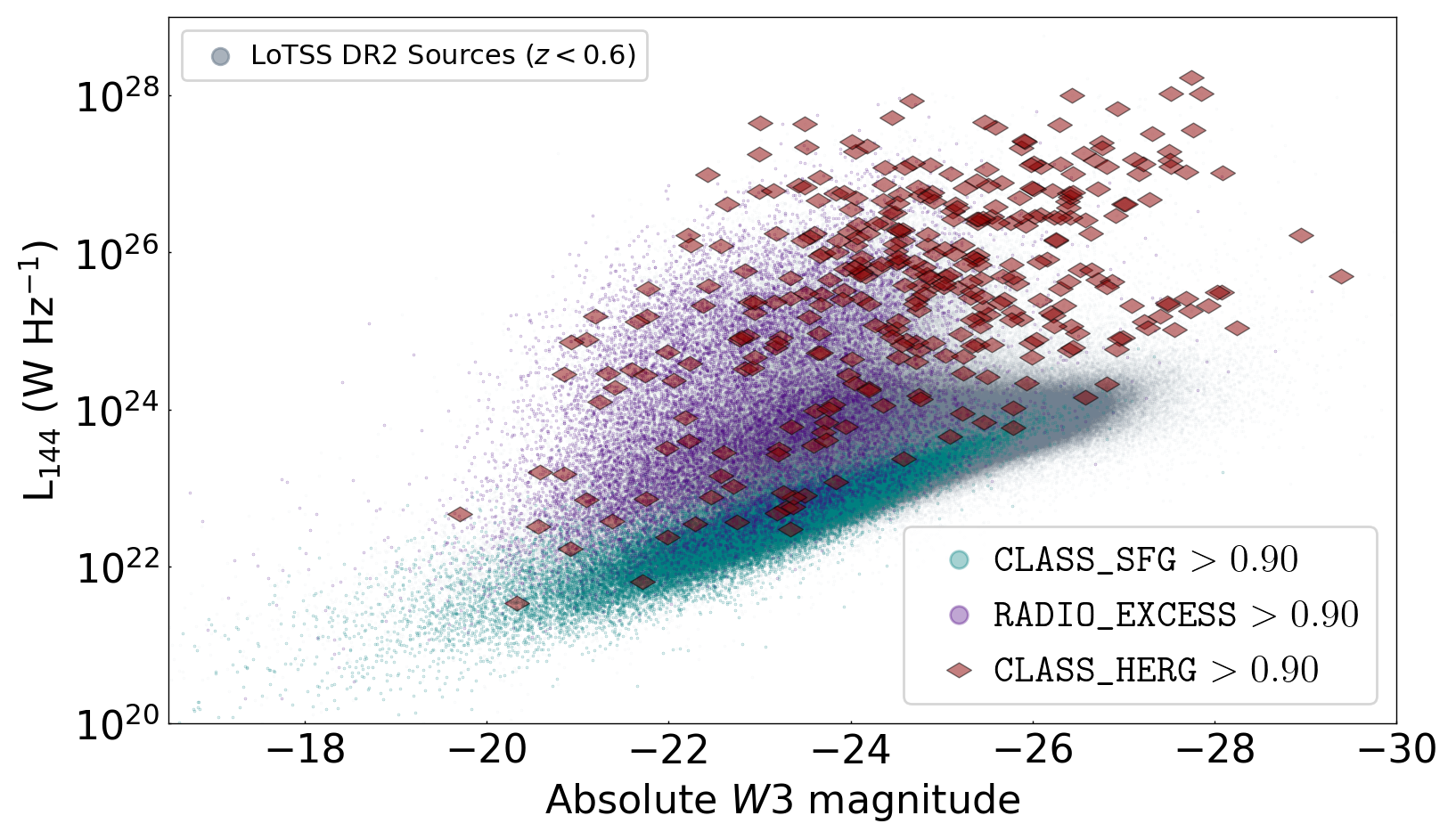}
    \caption{Radio luminosity as a function of (left) $W2$ and (right) $W3$ absolute magnitude. The logarithmic density plot in grey shows 1.0 million DR2 sources with $z<0.6$ as a comparison population. Colours as in Fig.\ \ref{fig:wisecc}.}
    \label{fig:wiseabs}
\end{figure*}

\begin{figure}
\includegraphics[width=0.48\textwidth]{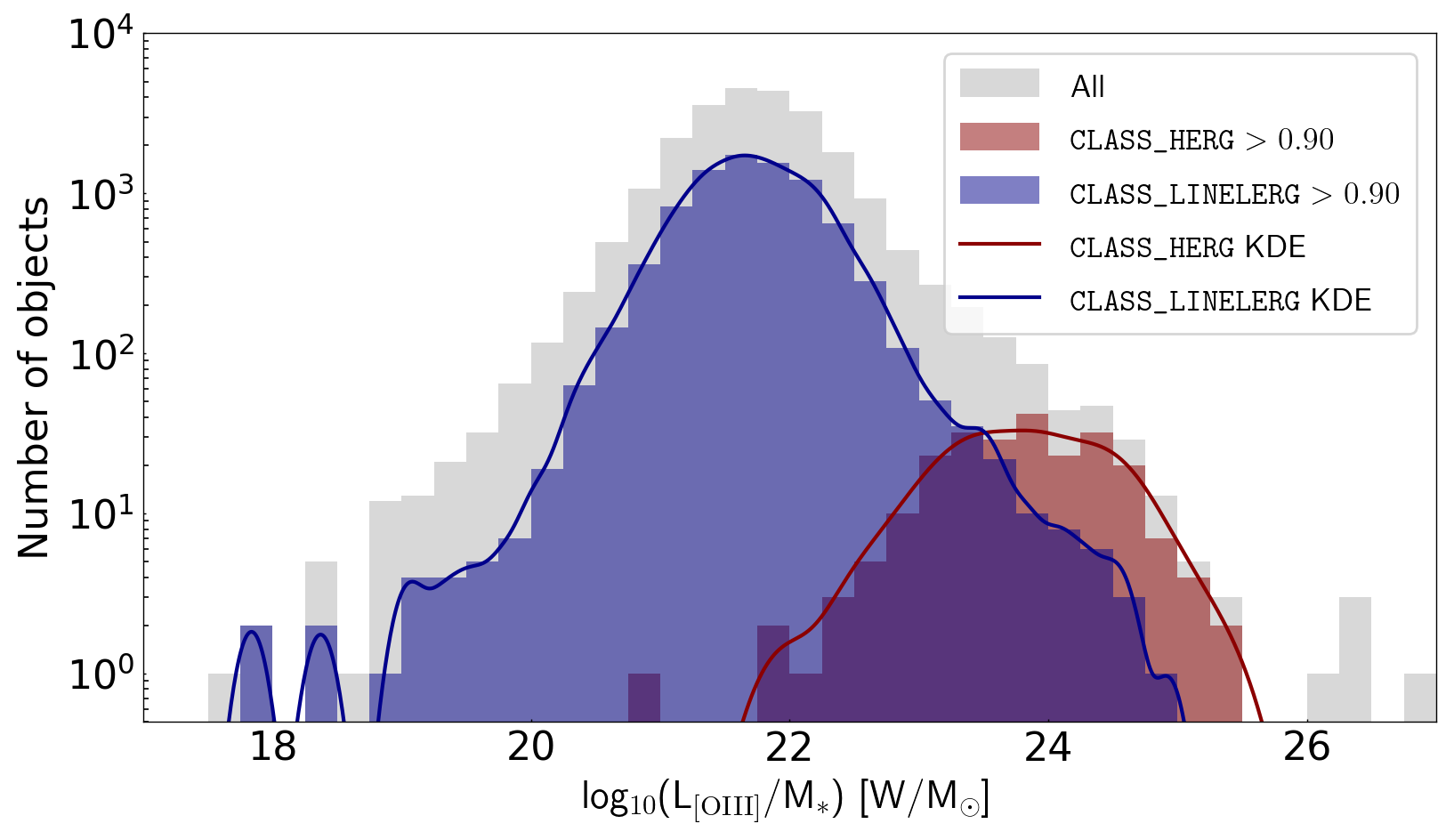}
\caption{The distribution of the specific [\textsc{Oiii}] luminosity for 19,331 radio-excess objects with mass estimates and [\textsc{Oiii}] measurements in our sample, colour coded by their emission-line classification status, if any. Histograms show the raw source counts, solid lines show kernel density estimates smoothing over the observed distribution.}
\label{fig:oiiihist}
\end{figure}

\subsection{Physical Properties of source by classification}

A subset of our sample was included in the value-added catalogue constructed by the MPA-JHU group \citep{Brinchmann2004}. While we chose to classify our objects using the emission line measurements presented  in the Portsmouth catalogue (which supersede the MPA-JHU measurements), we can use the MPA-JHU physical parameters (e.g. SFR and stellar mass; M$_{\rm{*}}$) as independent validation of our classification technique. In Figure \ref{fig:sfgs_plus_contours} for instance we show two panels to demonstrate the distribution of our securely classified objects (applying a 90\% threshold) in the 
 log$_{\rm{10}}$(M$_{\rm{*}}$/M$_{\rm{\odot}}$) vs log$_{\rm{10}}$SFR, and  log$_{\rm{10}}$SFR vs log$_{\rm{10}}$L$_{\rm{144}}$ planes. In the left-hand panel we see that the objects with a reliable classification as {\texttt{CLASS\_SFG}} form a ``main sequence" of SFGs in  log(M$_{\rm{*}}$/M$_{\rm{\odot}}$) vs log$_{\rm{10}}$SFR, which aligns well with the parameterisation from \citet[][evaluated at $z = 0.2$ and converted to the initial mass function from \citealt{Kroupa2001}, which was adopted for the MPA-JHU catalogue]{Schreiber2015} which  has been overlaid as the black line with a shaded region indicating the typical $\pm 0.3$\,dex scatter \citep[e.g.][]{Tacchella2016}. All three classes of AGN; RQAGN, HERG and LINELERG scatter in large part below the main sequence, with some level of overlap. Our source distributions demonstrate that our probabilistic spectral source classifications broadly select objects with the expected range of stellar masses and SFRs for their physical class, measured through an independent catalogue. The distribution is also qualitatively similar to that seen in \cite{Gurkan2018} who used early LOFAR observations of the H-ATLAS NGP to classify $\approx 15\,000$ radio sources in a search for SFGs. By using \cite{Best2012} spectral classifications to isolate radio-AGN, and in conjunction with a BPT analysis applied to MPA-JHU emission line measurements \citep{Brinchmann2004}, the authors verified their SED-fitting--derived SFR measurments by showing that the classified sources fall as expected relative to the main sequence of SFGs (which in turn is well-aligned with the \citealt{Elbaz2007} main-sequence relation). In the right-hand panel of Figure \ref{fig:sfgs_plus_contours}, we show the log$_{\rm{10}}$SFR vs log$_{\rm{10}}$(L$_{\rm{144}}$) plane, combining the LOFAR log$_{\rm{10}}$(L$_{\rm{144}}$) measurement \citep{Hardcastle2023} with the MPA-JHU SFR, and colour-coding by our independently-measured source classifications. Our securely-classified SFGs (90\% threshold) form a sequence in L$_{\rm{144}}$ -- log$_{\rm{10}}$SFR -- the well-known radio-SFR relation (e.g. \citealt{Smith2021}, which has been overlaid as the black dot-dashed line), whereas securely-classified HERGs and LERGs scatter above the relation as expected, due to the excess in their radio luminosities as a consequence of their AGN activity. It is worth re-iterating once again, that, although the HERG and LERG classes are indeed selected on the basis of a radio excess relative to the L$_{\rm{144}}$ predicted by their SFR (in our case their Balmer-corrected H$\alpha$ luminosity which directly correlates with SFR) the values of SFR in Figure \ref{fig:sfgs_plus_contours} have been derived independently in the MPA-JHU catalogue \citep{Brinchmann2004}, and thus serve as external validation of our analysis. Of particular interest in this panel are the RQAGN. These objects largely overlap with the main body of SFGs, but exhibit a tail of objects towards low stellar masses (log$_{\rm{10}}$SFR$\lessapprox -1$) with radio luminosities elevated slightly above the radio-SFR relation. The nature of the radio emission in these objects is likely to be studied with increasing levels of interest as radio surveys continue to progress; for instance as \texttt{LOFAR2.0} comes online (and in a number of pioneering works harnessing the international long baselines of LOFAR already; e.g. \citealt{Morabito2022}, \citealt{Sweijen2023}) imaging at 144 MHz will resolve $\sim$ kpc scales, potentially revealing the presence of either small-scale radio jets, or distributed, highly-obscured star formation.

\subsection{Additional tests}

\subsubsection{WISE colour and luminosity validation} 

The {\it WISE} colour-colour plot is a widely used diagnostic for
separating different classes of galaxies \citep[e.g.][]{Mateos+12,Gurkan14}. In Fig.\ \ref{fig:wisecc} we show the colour-colour plot for galaxies in
the DR2 optically identified sample \citep{Hardcastle2023} with galaxies classified at high confidence levels from the present paper
overlaid. It can be seen that star-forming galaxies and radio-excess objects both generally lie on the normal galaxy continuum with $0< W1-W2 <0.5$ (i.e. near-infrared colours dominated by an old stellar population) but are clearly distinguished by their $W2-W3$ colours, with SFGs having much redder $W2-W3$ due to the importance of warm dust in their overall SEDs. HERGs clearly lie away from the bulk of the (presumably LERG) radio-excess population, which is expected due to the contribution of the AGN torus to $W3$, and was seen in the similar analysis of bright radio sources by \cite{Gurkan14}, although there the proportion of HERGs was far larger. Also as expected, some of our HERGs lie in the location generally populated by quasars and likely could be classified as quasars, Seyfert 1s or broad-line radio galaxies. Importantly, many of the HERGs lie in the region generally populated by SFGs, showing that it is not safe to separate SFGs and radio-excess AGN by their WISE colours alone.

\begin{figure*}
	\includegraphics[width=\textwidth]
        {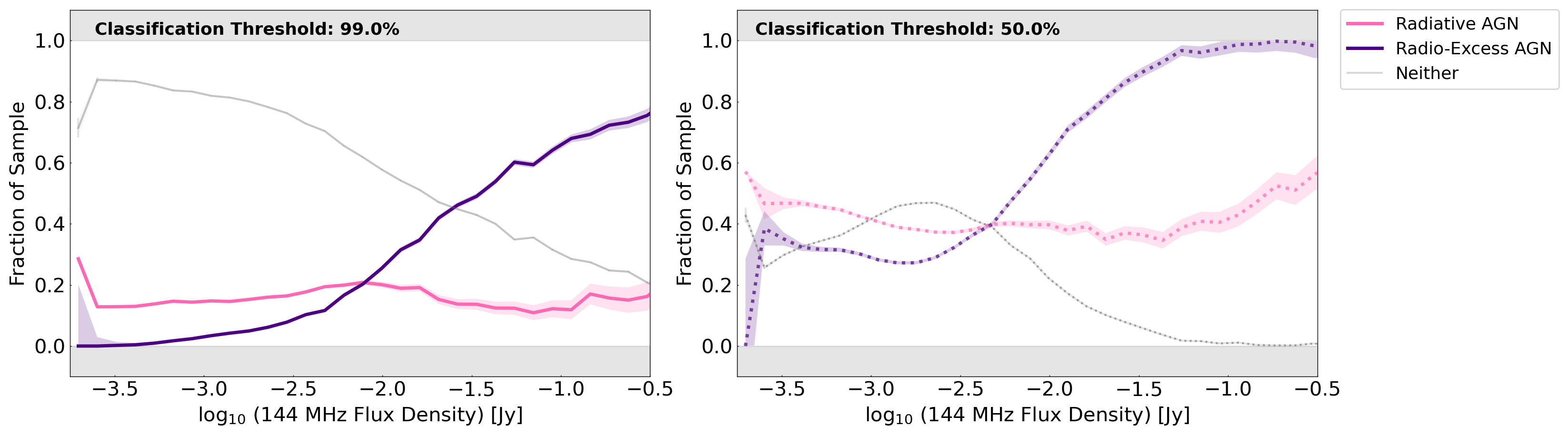}
        \includegraphics[width=\textwidth]{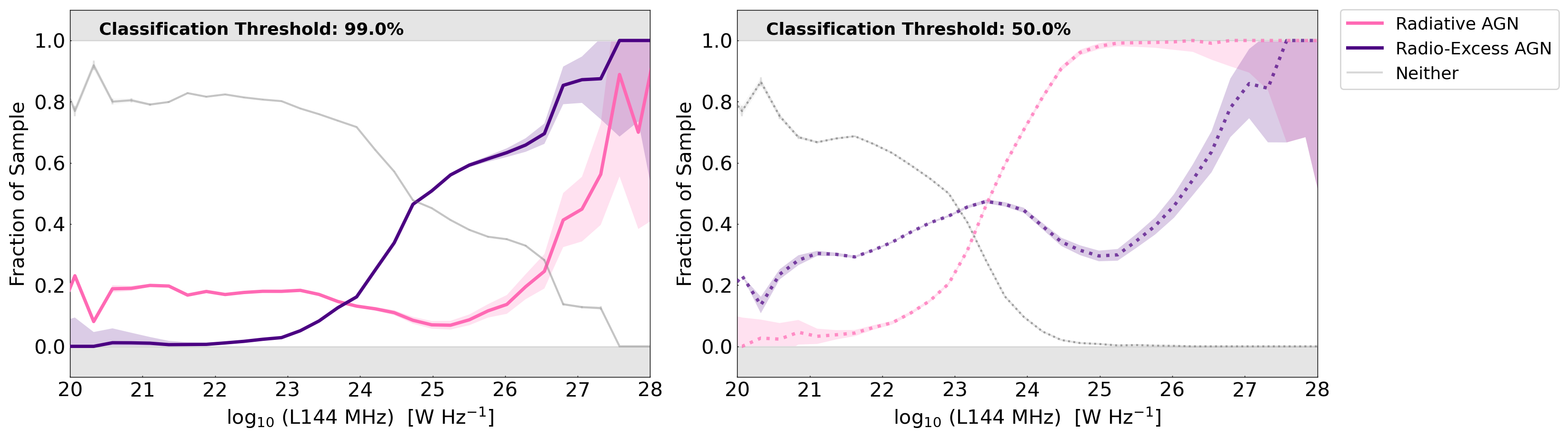}
    \caption{The fraction of sources identified as having a $>99$ ($>50$)\,percent reliability of hosting a radiative AGN, having a radio excess, or neither, are shown as pink, purple and grey, respectively (as indicated in the legend) as a function of 144\,MHz flux density (top) and luminosity (bottom panel). Error bars have been calculated assuming Poisson statistics. }
    \label{fig:demog_rad_rad_halves}
\end{figure*}

As shown by \cite{Hardcastle2023} we can also relate the {\it WISE} luminosity (absolute magnitude, here computed using a simple power-law extrapolation between {\it WISE} bands) to the radio luminosity for these sources, and this is shown in Fig. \ref{fig:wiseabs}. Another validation of our methods is that objects that we classify as SFGs or radio-excess objects are very differently positioned on these plots, and behave in different ways for $W2$ and $W3$ absolute magnitudes. $W2$ acts as a proxy of galaxy mass, and we see that radio-excess objects start to appear only in massive galaxies: a $W2$-$L_{144}$ correlation appears for SFG because of the relation between galaxy mass and star-formation rate. However, there is a clear region of overlap between SFG and radio-excess objects. By contrast the two populations are very well separated on the $W3$ plot where the correlation arises because of the relationship between dust luminosity and star-formation rate that gives rise to the well-known (far)-infrared/radio correlation \citep[e.g.][]{vanderkruit1971,Yun2001,Jarvis2010,Smith2014}. Radio-excess objects lie significantly above the main sequence of star formation as indicated by this correlation. Our emission-line classifications would allow us to calibrate a method of selecting radio-excess objects using only their {\it WISE} and radio properties, in a similar way to the {\it Herschel}-based approach of \cite{Hardcastle2016}. On these plots HERGs tend to be found at high radio luminosities, but also at high $W3$ absolute magnitudes, as noted by \cite{Gurkan14}.

\subsubsection{Building on this framework: Specific [\textsc{Oiii}] luminosity of HERGs \&\ LERGs validation}
\label{sec:martin}

In the picture in which the HERG/LERG dichotomy is a transition between radiatively efficient and inefficient AGN activity occurring at a level of a few per cent of the Eddington rate \citep{Best2012,Mingo2014,Hardcastle18}, we expect proxies of $L/L_{\rm Edd}$ to show a transition in samples that have reliable optical emission-line classifications. Locating this transition could also help us to classify objects that do not exhibit all the lines needed for a full BPT classification using our method.

In radio-excess objects, we can take the luminosity in the [\textsc{Oiii}] line as a proxy for the radiative activity of the AGN; standard linear corrections exist to turn this into an approximate bolometric luminosity \citep{Heckman2004} although these of course only apply to radiatively efficient objects. We do not have the black hole mass estimates needed to estimate the Eddington luminosity, but can make use of the galaxy mass/black hole mass relation \citep[e.g.][]{Reines2015} as a proxy of $M_{\rm BH}$. In this analysis we use the galaxy mass estimates from \cite{Hardcastle2023}, which are derived from the DESI Legacy Survey and WISE photometry and so are superior to those based on SDSS photometry, but are not available for all sources. 

Fig.\ \ref{fig:oiiihist} shows a histogram of the specific [\textsc{Oiii}] luminosity for radio-excess objects ($p>0.95$) broken down into HERG and LINELERG classes (where we use $p=0.8$ as the threshold for the emission-line classification). As expected (though not universally seen: \citealt{Whittam2022}), the objects we classify as HERGs have systematically high specific [\textsc{Oiii}] luminosity, although there is substantial overlap, as seen by \cite{Best2012} and Kondapally et al. (\textit{in prep}), which is not surprising as many of the LERG are not significantly detected in this line. Moreover, the distribution for HERGs is at least centred in roughly the right location. Take the critical accretion rate for radiatively efficient accretion to be $0.03{\dot M}_{\rm Edd}$, and we know that ${\dot M}_{\rm Edd}$ = $2.3 \times 10^{-9} M_{\rm BH}$ yr$^{-1}$. The black hole mass to galaxy mass ratio is around 0.003 for elliptical galaxies \citep{KormendyHo2013,Reines2015}, thus the critical accretion rate per galaxy mass is $\sim 2 \times 10^{-13}$ yr$^{-1}$, or $6.5 \times 10^{-21}$ s$^{-1}$. For an AGN efficency of $\eta=0.1$ and a bolometric correction of $B=3500$ from \cite{Heckman2004}, we expect the HERG/LERG transition at $6.5 \times 10^{-21} \eta M_{\odot} c^2/B = 3.3 \times 10^{22}$ W M$_\odot^{-1}$, which is in line with what is observed. The HERG distribution should in theory only be 1.5 -- 2 decades wide, but is of course broadened by the decade or so of intrinsic scatter in the black hole mass/galaxy mass relation, as well as uncertainties on efficiency factors, bolometric corrections and the measurements of galaxy mass and luminosity themselves.

Fitting a KDE estimator to the histogram allows us to see that a naive Bayesian classifier would state that any object with $\log_{10}(L/M*) > 23.5$ is most likely to be a HERG, and applying this to the unclassified sources in our dataset, which typically do not have all the lines detected that are needed for a BPT classification, would allow us to select an additional $\sim 80$ objects that could be classified as HERGs. This modification to the scheme would effectively reinstate the `equivalent width of [\textsc{Oiii}]' criterion for HERG selection used in some previous work \citep{Laing1994}, but now with an explicit physical motivation. Given the small numbers involved, we do not attempt to apply this classification to the present catalogue. The nature of the HERG and LERG in these samples, making use of this method, and the relationship between emission-line class and radio properties will be investigated in a future paper (Chilufya et al in prep).

\begin{figure*}
	\includegraphics[width=\textwidth]{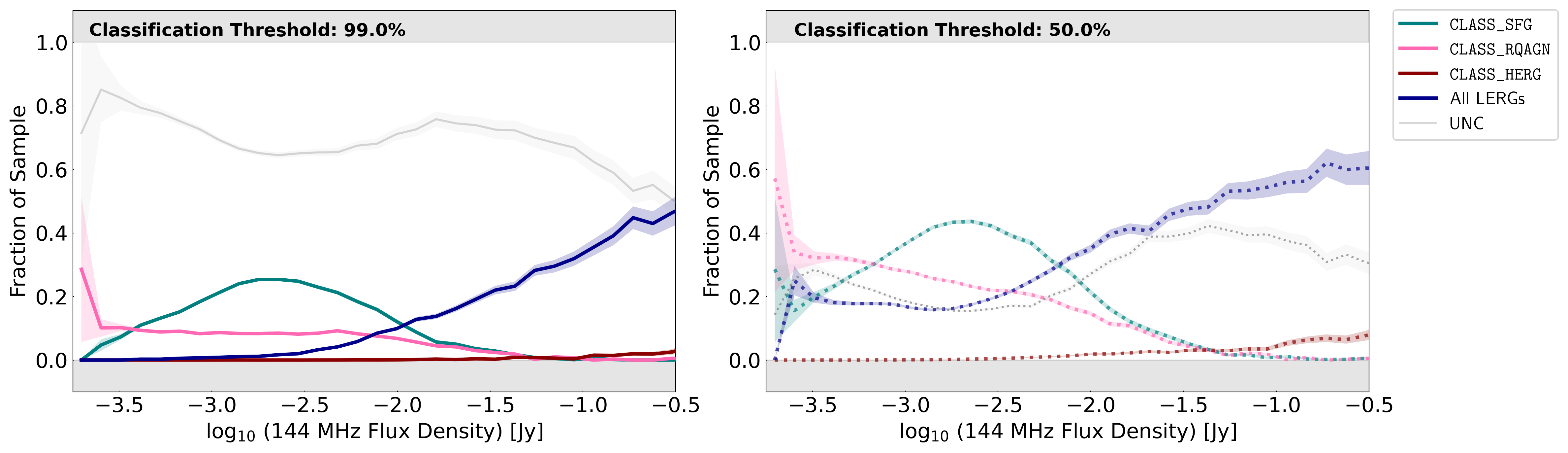}
        \includegraphics[width=\textwidth]{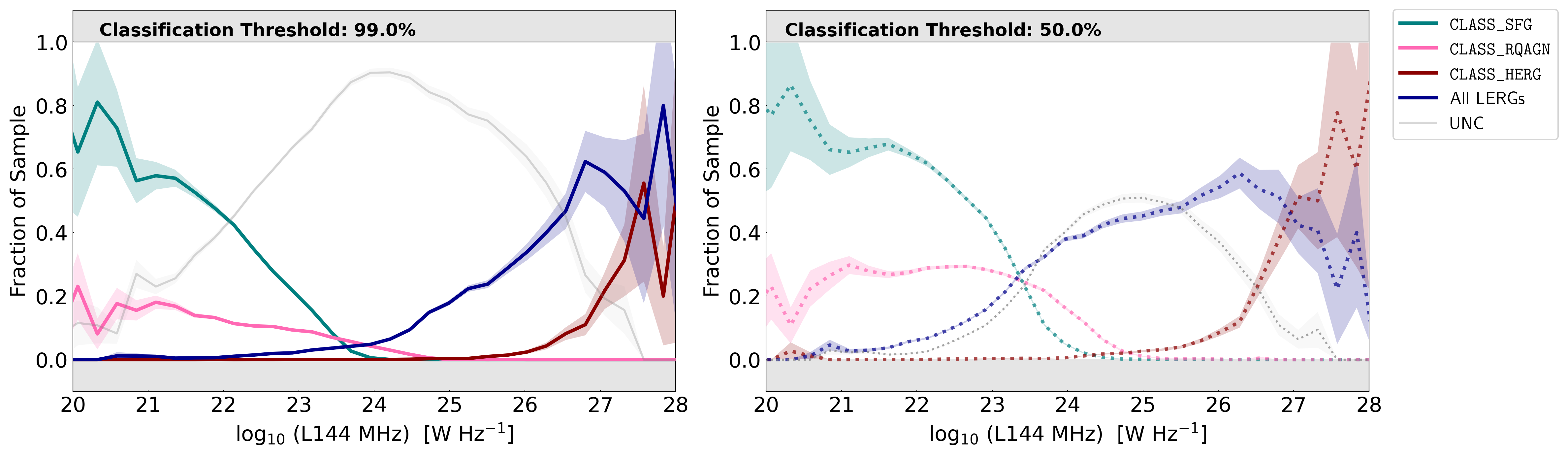}
        \includegraphics[width=\textwidth]{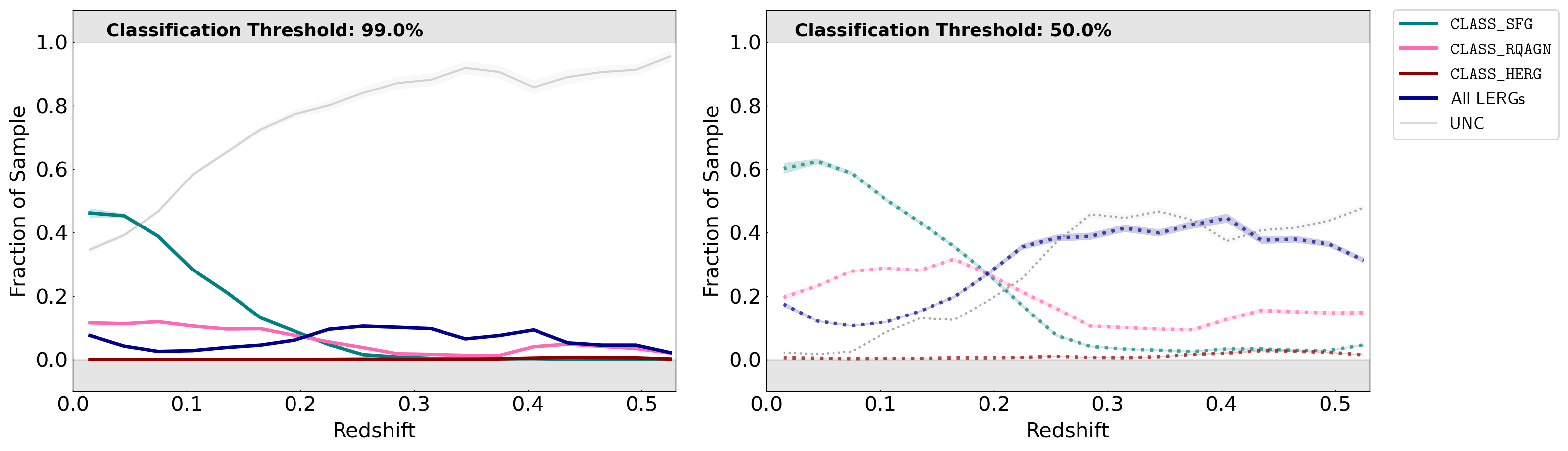}
    \caption{The fraction of sources with $>99$($>50$)\,percent reliability classifications as SFG, RQAGN, HERGs, LINELERGs are shown as teal, pink, red and blue symbols, respectively (as indicated in the legend) as a function of 144\,MHz flux density (top), 144\,MHz luminosity (middle) and redshift (bottom). In each panel, sources that do not reach the classification threshold in any of the four categories are shown as unclassified ("UNC", and indicated in grey). Error bars have been estimated using Poisson statistics.}
    \label{fig:demog_quads}
\end{figure*}

\section{Spectroscopic Source demographics in LoTSS DR2}
\label{sec:results}

\subsection{Overall Source Demographics}
\label{sec:demographics}

We describe in this section the demographic breakdown of the LoTSS DR2 sources which have been studied in this work. As we discuss in Section \ref{subsect:probabilistic_classif} one can construct a particular physical class of objects using different reliability criteria according to the need for purity or completeness of a given sample. This is especially straight-forward for the physical-class reliabilities already provided in our catalogue, but we can also utilise our probabilistic \radioexcess\ measurements, and re-construct a class of radiative AGN (i.e. containing objects in either \classrq\ or \classherg) to represent numbers along the two independent axes with which we identify AGN. As a demonstration of two potential approaches to purity and completeness, we conduct the following demographic analysis applying two different thresholds - a very high reliability threshold ($>99$\% reliability) and a very permissive threshold ($>50$\% reliability) requiring that objects must be classified into the same class for at least 99\%/50\% of the catalogue realisations. \\

\noindent Initially we examine the breakdown of the subsets of objects likely to be radiative- and radio-excess AGN as a function of radio flux density and radio luminosity. These subsets are constructed by selecting objects with {\texttt{BPT\_SFG}} $<0.01$ ($0.50$) to capture all subclasses of radiative AGN, and by applying a reliability cut of $0.99$ ($0.50$) in \radioexcess\ respectively. These breakdowns are shown in Figure \ref{fig:demog_rad_rad_halves}. In the top left-hand panel we show our most reliable measurements of the two diagnostics as a function of flux density (requiring that $>99\%$ of realisations satisfy each criterion). Below flux levels of $\approx25$ mJy the dominant population consists of objects which do not satisfy either of these criteria 99\% of the time, whereas at the highest fluxes $>25\,$mJy, radio-excess sources dominate the source counts. Radiative AGN make up roughly 20\% of the population at all flux levels. In the top right-hand panel we show the equivalent plot this time applying a liberal $50$\% threshold. In this case we see radio excess objects constituting approximately 30\% of our sample up to a flux level of $\approx 3$ mJy, before rising smoothly to make up almost 100\% of the sample above flux densities $\approx 40$ mJy. 

In the lower-left panel of Figure \ref{fig:demog_rad_rad_halves}, as a function of log luminosity, we see similar behaviour in that below log$_{\rm10}\,$(L$_{\rm{144}}\,$/$\,$W$\,$Hz$^{-1}$) $\approx$ 24.5, the majority of sources are neither radiative AGN or Radio-Excess AGN in 99\% of realisations. Again the radiative AGN compose an approximately constant 20\% of sources below log$_{\rm10}\,$(L$_{\rm{144}}\,$/$\,$W$\,$Hz$^{-1}$) $\approx$ 24.5, they then drop slightly before making a steady rise to $\approx 80$\% by log$_{\rm10}\,$(L$_{\rm{144}}\,$/$\,$W$\,$Hz$^{-1}$) $\approx$ 27. The radio excess AGN rise steeply from zero beginning at log$_{\rm10}\,$(L$_{\rm{144}}\,$/$\,$W$\,$Hz$^{-1}$) $\approx$ 23, reaching approximately 100\% by log$_{\rm10}\,$(L$_{\rm{144}}\,$/$\,$W$\,$Hz$^{-1}$) $\approx$ 27.5. This reflects the behaviour we expect for radio-selected samples, as we discuss in Section \ref{subsect:further_tests}. In the lower right panel we apply our liberal $50$\% threshold, finding that only below log$_{\rm10}\,$(L$_{\rm{144}}\,$/$\,$W$\,$Hz$^{-1}$) $\approx$ 23 do sources which are neither radiative AGN or radio-excess AGN dominate the sample. At approximately the same luminosity radiative AGN begin to dominate the fractional counts, rising steeply to $\approx 100$\% by log$_{\rm10}\,$(L$_{\rm{144}}\,$/$\,$W$\,$Hz$^{-1}$) $\approx$ 24.5. Radio excess object also reach $\approx 100$\% but only at the higher luminosity of log$_{\rm10}\,$(L$_{\rm{144}}\,$/$\,$W$\,$Hz$^{-1}$) $\approx$ 27.5. \\


\noindent We next consider four non-overlapping physical classes of sources, classified as described mainly in Section \ref{Sect: Method}. In addition to \classsfg, \classrq, and \classherg, we combine our recorded diagnostics to define the final physical class of objects: LERGs (low-excitation radio galaxies). To reconstruct the full LERG class requires that we also consider the information gained from the \emph{lack} of measured emission lines. As such, we combine the emission-line LERG class (\texttt{CLASS\_LINELERG}) with those objects with a \radioexcess\ reliability greater than our chosen threshold, and that show a \classherg\ reliability less than 1 minus this threshold (i.e. we want to select the sources least likely to lie in \classherg, as by definition LERGs are radio-excess sources which are not highly ionised). \\

\noindent In Figure \ref{fig:demog_quads} we show three rows of panels which record the demographic break down of securely-classified ($>99$\% reliability; left-hand column) and ``liberally"-classified ($>50$\% reliability; right-hand column) sources in each of the four physical classes described above. Objects which do not meet the threshold in any one class are recorded as unclassified ("UNC"). \\ 

In the upper panels we show the demographics as a function of the radio flux at 144 MHz; log$_{\rm{10}}$(F$_{\rm{144}}$). In the left-hand panel, as a result of our strict criterion to ensure high purity of physical classes, the majority of sources appear in the UNC subset at all flux densities. Amongst securely-classified subsets the LERGs dominate above $\approx\!10\,$mJy, reaching $\approx\!50$\% of sources at the highest flux densities ($\approx 315\,$mJy). Below $\approx\!10\,$mJy SFGs are the dominant population, peaking at $\approx\!2.2\,$mJy. RQAGN make up approximately $10$\% of the population below $\approx\!10\,$mJy (the same threshold at which LERGs become more numerous than SFGs) and gradually decline above this, reaching roughly zero by $\approx\!40\,$mJy. The HERG class are barely visible on this plot, reaching only a few percent of the fractional counts by flux densities above $\approx\!100\,$mJy. In the right-hand panel where we apply a liberal threshold of $50$\% very similar behaviour is seen across all physical classes, only with fractional counts scaled up by about $10$\%. This results in the unclassified class "UNC" dropping substantially to lie below the individual physical classes down to around $\approx\!1\,$mJy, more in-line with previously reported demographics, where the doubt on any individual classification has not been quantified.\\

In the central panels of Figure \ref{fig:demog_quads} we present the demographic breakdown of sources as a function of luminosity, plotted as log$_{\rm10}\,$(L$_{\rm{144}}\,$/$\,$W$\,$Hz$^{-1}$). In the left-hand panel showing our most securely-classified sources, it is interesting to note that across the entirety of the central luminosity range; \mbox{$22.0 <$ log$_{\rm10}\,$(L$_{\rm{144}}\,$/$\,$W$\,$Hz$^{-1}$) $< 26.5$;} objects with no secure classification dominate the fractional counts, whereas at the lowest luminosities i.e.log$_{\rm10}\,$(L$_{\rm{144}}\,$/$\,$W$\,$Hz$^{-1}$)$\,< 22.0$, SFGs dominate the source counts. On the right of the plot, at the highest luminosities, LERGs and HERGs dominate the source counts, exceeding unclassified objects at log$_{\rm10}\,$(L$_{\rm{144}}\,$/$\,$W$\,$Hz$^{-1}$)$\,\approx\!26.25$ and log$_{\rm10}\,$(L$_{\rm{144}}\,$/$\,$W$\,$Hz$^{-1}$)$\,\approx\!27.0$ respectively. This panel in particular goes to show the importance of incorporating a `quantified doubt' when reporting demographics, or examining the behaviour of a particular class of objects - particularly across the luminosity range where it happens that most LoTSS-Wide detections lie. In the right-hand panel we again see behaviour of the physical classes very similar to the most secure classifications -- for SFGs and RQAGN the fractional counts again seem to simply scale up by about $10$\%, meanwhile the LERGs and HERGs are slightly more complex. Unclassified objects drop dramatically to being the dominant class only for $\approx\!2$ decades, and are only very slightly ($<10$\%) more numerous than the LERGs in this range. The LERG class interestingly receives a boost across all luminosities up to about log$_{\rm10}\,$(L$_{\rm{144}}\,$/$\,$W$\,$Hz$^{-1}$)$\,\approx\!26$ whereby their counts actually drop more so than for the securely-classified subset, and appear to shift into the HERG class instead. This is a potentially interesting observation if this ambiguity between the HERG and LERG classes is uncaptured in other works. \\

Finally, in the lower panels of Figure \ref{fig:demog_quads} we show our demographic breakdown as a function of spectroscopic redshift, $z$ (as recorded in the Portsmouth catalogue). While it is interesting to examine these numbers, our combination of spectroscopic catalogues that target different classes of objects and across two different redshift ranges (as well the fact that our sample comprises only a few percent of LoTSS sources) means the demographics as a function of redshift are unlikely to be representative of the full population of galaxies in the Universe. 

For our most securely classified sources (in the left-hand panel) only at the very lowest redshifts does any one physical class (the SFGs) dominate above unclassified sources. At $z > 0.05$, unclassified sources are by far the most numerous class, reaching 75\% of the sample by $z \approx 0.15$, rising almost to 100\% by the highest redshifts in this work ($z \approx 0.55$). Considering only the securely classified subsets, SFGs drop rapidly, reaching zero by $z \approx 0.255$. RQAGN constitute $\approx$ 17\% of the population out to $z \approx 0.2$, with a secondary bump at $z = 0.4$ -- 0.5. LERGs are at very low levels across most redshifts, but reach $\approx$ 17\% across the $z = 0.25$ -- 0.4 range. HERGs are likewise at very low ($<$10\%) levels across the entire redshift range. In the right-hand panel which shows our more liberally selected subsets, we find unclassified sources drop to a level comparable with LERGs across the majority of redshifts ($z \gtrapprox\!0.2$). The LERGs are once again the most dramatically boosted class by lowering the reliability threshold, meanwhile SFGs and RQAGN scale upwards by $\approx\!10$\%, and while this may also be true of the HERGs, they remain undetectable in the fractional counts until a rise at $z \gtrapprox\!0.35$ to a few percent of the total counts. \\

\section{Conclusions}
\label{Sect: Concl}

We have produced probabilistic classifications for the $\sim 4$ per cent of LoTSS DR2 \citep{Shimwell2022} sources for which spectroscopic measurements exist in the Portsmouth catalogue \citep{Thomas2013}, based on their analysis of SDSS and BOSS spectra. 
We have done this by combining the spectroscopic and low-frequency radio information with extensive Monte-Carlo simulations, in which we have produced multiple realisations of the input catalogues in a manner that accounts for the measured uncertainties.
Since our classifications are probabilistic, they can be used to produce samples of sources tailored according to the scientific requirements of the end user, whether that be high purity (e.g. including only the most reliably-classified High-Excitation Radio Galaxies) or completeness (e.g. including any source that may have a radio excess). 

Using a 90 percent reliability threshold, in the LoTSS DR2 sample we have identified \Nradexs\ radio-excess AGN (i.e. those which have an excess of 144\,MHz emission over that which would be expected on the basis of their Balmer-corrected H$\alpha$ line alone), and identified \NLINELERG\ LINELERGs and \NHERG\ HERGs among them. Similarly, we have been able to identify \NSFG\ SFGs and \NRQAGN\ RQAGN. Our catalogue contains the largest sample of spectroscopically-classified radio sources to date, far surpassing previous works in the literature \citep[e.g.][]{Best2012,Sabater2019,Whittam2022}.

We have validated our method by comparing our results with the wider literature, including comparisons with previous catalogues such as \citet{Best2012}, with other diagnostics such as the Excitation Index \citep{Buttiglione2010}, and with our expectations based on our previous investigations of the radio source population \citep[e.g.][]{Best2023, Das2024}.

Although our method has produced classifications for only around 4 per cent of the LoTSS DR2 sources, our method has been designed to be directly applicable to the observations of the WEAVE-LOFAR survey. WEAVE-LOFAR \citep{Smith2016} will -- in 2024 -- begin to produce more than a million spectra of sources identified purely on the basis of their 144\,MHz emission, and include spectroscopy of every radio source detected in the LoTSS deep fields \citep{Tasse2021,Sabater2021,Kondapally2021,Duncan2021}.

\section*{Acknowledgements}

ABD, DJBS, MJH and MIA acknowledge support from the UK Science and Technology Facilities Council (STFC) grant ST/V000624/1. DJBS also acknowledges STFC support from grant ST/Y001028/1. PNB is grateful for support from the UK STFC via grant ST/V000594/1. RK is grateful for support from the UK STFC via grant ST/V000594/1. MIA acknowledges support from the UK Science and Technology Facilities Council (STFC) studentship under the grant ST/V506709/1. SD acknowledges support from a Science and Technology Facilities Council (STFC) studentship via grant ST/W507490/1. SS acknowledges support from the UK Science and Technology Facilities Council (STFC) via grant ST/X508408/1. KJD acknowledges support from the STFC through an Ernest Rutherford Fellowship (grant number ST/W003120/1). For the purpose of open access, the author has applied a Creative Commons Attribution (CC BY) licence to any Author Accepted Manuscript version arising from this submission.\\

LOFAR is the Low Frequency Array, designed and constructed by ASTRON. It has observing, data processing, and data storage facilities in several countries, which are owned by various parties (each with their own funding sources), and which are collectively operated by the ILT foundation under a joint scientific policy. The ILT resources have benefited from the following recent major funding sources: CNRS-INSU, Observatoire de Paris and Université d'Orléans, France; BMBF, MIWF-NRW, MPG, Germany; Science Foundation Ireland (SFI), Department of Business, Enterprise and Innovation (DBEI), Ireland; NWO, The Netherlands; The Science and Technology Facilities Council, UK; Ministry of Science and Higher Education, Poland; The Istituto Nazionale di Astrofisica (INAF), Italy.

This research made use of the Dutch national e-infrastructure with support of the SURF Cooperative (e-infra 180169) and the LOFAR e-infra group. The Jülich LOFAR Long Term Archive and the German LOFAR network are both coordinated and operated by the Jülich Supercomputing Centre (JSC), and computing resources on the supercomputer JUWELS at JSC were provided by the Gauss Centre for Supercomputing e.V. (grant CHTB00) through the John von Neumann Institute for Computing (NIC).

This research made use of the University of Hertfordshire
high-performance computing facility and the LOFAR-UK computing
facility located at the University of Hertfordshire (\url{https://uhhpc.herts.ac.uk}) and supported by
STFC [ST/P000096/1], and of the Italian LOFAR IT computing
infrastructure supported and operated by INAF, and by the Physics
Department of Turin University (under an agreement with Consorzio
Interuniversitario per la Fisica Spaziale) at the C3S Supercomputing
Centre, Italy.

\section*{Data Availability}

This paper is based on LoTSS DR2 optical identifications. The optical identification catalogue, together with the classifications produced in this work (and the documentation necessary to use them) are available to download from \url{https://lofar-surveys.org/dr2_release.html}. Other data presented in the paper are available upon reasonable request to the corresponding author.



\bibliographystyle{mnras}
\bibliography{example,additional_refs} 




\appendix

\section{Example Table and Subsets}


\begin{table*}
    \centering
            \caption{An example of the machine-readable data table we provide with this work. Columns are described in Table \ref{tab:col_head}. Four objects are included, two for which BPT analysis and classifications have been derived (objects 1 and 2), and two for which radio-excess analysis was carried out, but BPT lines were not available for further analysis (and hence full classification; objects 3 and 4).}
    \begin{tabular}{lllll}
    \hline
    \hline
\bf{{Column}} & \bf{{Object1}} & \bf{{Object2}} & \bf{{Object3}} & \bf{{Object4}} \\ 
\hline
\hline
{\bf{\texttt{Source$\_$Name}}} & ILTJ000001.80$+$193250.3 & ILTJ000002.57$+$200308.5 & ILTJ000004.53$+$220436.5 & ILTJ000007.23$+$332358.6 \\ 
 {\bf{\texttt{SPECOBJID}}} & 6946961409339748352 & 6898653816788021248 & 6879497850414886912 & 8043642899621670912 \\ 
 {\bf{\texttt{TOT$\_$F144(mJy)}}} & 3.04097 & 3.03803 & 5.60489 & 3.21012 \\ 
 {\bf{\texttt{ID$\_$RA}}} & 0.00738 & 0.01082 & 0.01880 & 0.03024 \\ 
 {\bf{\texttt{ID$\_$DEC}}} & 19.54727 & 20.05232 & 22.07691 & 33.39955 \\ 
 {\bf{\texttt{z$\_$best}}} & 0.09049 & 0.27949 & 0.47511 & 0.48936 \\ 
 {\bf{\texttt{REDSHIFT}}} & 0.09067 & 0.27973 & 0.47530 & 0.48958 \\ 
 {\bf{\texttt{CLASS$\_$z$\_$WARNING}}} & 0.000 & 0.000 & 0.000 & 0.000 \\ 
 {\bf{\texttt{RADIO$\_$EXCESS}}} & 0.000 & 0.053 & 0.569 & 0.999 \\ 
 {\bf{\texttt{BALMER$\_$CORR$\_$WARNING}}} & 0.000 & 0.000 & 0.000 & 1.000 \\ 
 {\bf{\texttt{BPT$\_$SFG}}} & 0.333 & 0.355 & -- & -- \\ 
 {\bf{\texttt{BPT$\_$COMP}}} & 0.667 & 0.642 & -- & -- \\ 
 {\bf{\texttt{BPT$\_$SEY}}} & 0.090 & 0.027 & -- & -- \\ 
 {\bf{\texttt{BPT$\_$LIN}}} & 0.577 & 0.618 & -- & -- \\ 
 {\bf{\texttt{BPT$\_$CSEY}}} & 0.090 & 0.027 & -- & -- \\ 
 {\bf{\texttt{BPT$\_$CLIN}}} & 0.577 & 0.615 & -- & -- \\ 
  {\bf{\texttt{BPT$\_$ML}}} & Y & Y & -- & -- \\ 
 {\bf{\texttt{zscore}}} & 0.55777 & 0.50295 & -- & -- \\ 
 \hline
 {\bf{\texttt{CLASS$\_$SFG}}} & 0.333 & 0.350 & -- & -- \\ 
 {\bf{\texttt{CLASS$\_$RQAGN}}} & 0.667 & 0.597 & -- & -- \\ 
 {\bf{\texttt{CLASS$\_$HERG}}} & 0.000 & 0.000 & -- & -- \\ 
 {\bf{\texttt{CLASS$\_$LINELERG}}} & 0.000 & 0.053 & -- & -- \\ 

 \hline
    \end{tabular}
    \label{tab:cat_example}
\end{table*}

In Table \ref{tab:cat_example} we present the first few rows of our table to demonstrate how objects will appear. Objects 1 and 2 are fully classified sources, meanwhile objects 3 and 4 demonstrate the information available for sources that do not exhibit 4 BPT emission lines, but for which we have measured a radio excess. Non-overlapping subsets of the BPT diagram are as follows:
\texttt{BPT\_SFG} + \texttt{BPT\_SEY} + \texttt{BPT\_LIN} must $= 1$ \\

\noindent To reconstruct the subset of objects above the \cite{Kewley06} maximal starburst line: \texttt{AGN\_Kewley} = \texttt{BPT\_SEY} + \texttt{BPT\_LIN} - \texttt{BPT\_COMP} \\

\noindent For clarity, the subsets \texttt{BPT\_CSEY} and \texttt{BPT\_CLIN} are subsets of \texttt{BPT\_COMP}. 
\texttt{BPT\_CSEY} + \texttt{BPT\_CLIN} = \texttt{BPT\_COMP} \\

\noindent As discussed in the text, to reconstruct a LERG class which incorporates all objects with a radio excess that have not been classed as HERGs, we apply a threshold of choice e.g. 90\% and compute the following: \texttt{RADIO\_EXCESS}$>0.90$ \& \texttt{CLASS\_HERG}$<0.10$



\bsp	
\label{lastpage}
\end{document}